\documentclass[]{interact}

\usepackage{epstopdf}

\usepackage[authoryear, square]{natbib}
\bibpunct[, ]{(}{)}{;}{a}{}{,}
\renewcommand\bibfont{\fontsize{10}{12}\selectfont}

\usepackage[utf8]{inputenc}
\usepackage{graphicx}
\usepackage{graphics}
\usepackage{moreverb}
\usepackage{epsfig}
\usepackage{hangcaption}
\usepackage{txfonts}
\usepackage{palatino}
\usepackage{amsmath}
\usepackage{adjustbox}
\usepackage{booktabs}
\usepackage{multirow}
\usepackage{xcolor,soul}
\usepackage{enumitem}
\usepackage{float}
\usepackage{lscape}
\usepackage{longtable}
\usepackage{setspace}

\theoremstyle{plain}

\theoremstyle{definition}

\theoremstyle{remark}

\renewcommand{\caption}[1]{\singlespacing\hangcaption{#1}}

\begin{document}

\title{Learning Financial Networks with High-frequency Trade Data}
\author{
\name{
Kara Karpman\textsuperscript{a}\thanks{Department of Statistics and Data Science, Cornell University. Email: kjk233@cornell.edu}
,
Sumanta Basu\textsuperscript{b}\thanks{Department of Statistics and Data Science, Cornell University. Email: sumbose@cornell.edu} and 
David Easley\textsuperscript{c}\thanks{Department of Economics, Cornell University. Email: dae3@cornell.edu}
}
}
\date{ }
\maketitle

\begin{abstract}
Financial networks are typically estimated by applying standard time series analyses to price-based economic variables collected at low-frequency (e.g., daily or monthly stock returns or realized volatility).  These networks are used for risk monitoring and for studying information flows in financial markets. High-frequency intraday trade data sets may provide additional insights into network linkages by leveraging high-resolution information. However, such data sets pose significant modeling challenges due to their asynchronous nature, nonlinear dynamics, and nonstationarity. To tackle these challenges,  we estimate financial networks using random forests. The edges in our network are determined by using microstructure measures of one firm to forecast the sign of the change in a market measure (either realized volatility or returns kurtosis) of another firm. We first investigate the  evolution of network connectivity in the period leading up to the U.S. financial crisis of 2007-09. We find that the networks have the highest density in 2007, with high degree connectivity associated with Lehman Brothers in 2006. A second analysis into the nature of linkages among firms suggests that larger firms tend to offer better predictive power than  smaller firms, a finding qualitatively consistent with prior works in the market microstructure literature.
\end{abstract}

\begin{keywords}
market microstructure; high-frequency trading; random forests
\end{keywords}

\section{Introduction}
\label{sec:ML_HFT_introduction}

From both theoretical and practical perspectives, there is interest in estimating linkages among financial institutions using data.  Academicians seek to understand how information flows between firms, regulators aim to identify when and how risk spreads through the financial system, and financiers would like to know whether incorporating other firms' characteristics can improve their own trading algorithms.  The matter of how firms interact with one another can be represented mathematically as a network, with nodes corresponding to financial institutions and an edge between two nodes indicating that those firms are connected in some sense.  

From a data scientist's perspective, there are two key questions as to how best to measure these connections. First, it is important to understand which statistical methods are well-suited for the task of estimating linkages.  Second, one needs to decide what type of data to apply these methods on. Financial institutions generate a variety of data through their activities (e.g. trading volumes and stock prices) and it is not immediately apparent what kind of data are most informative for measuring firms' connectivity.

Predictability of one firm's future performance using another firm's past is often used to define edges between two firms in financial networks. There is a rich literature applying linear and vector autoregression (VAR) methods to the stochastic process of a firm's stock price, e.g., its stock returns or return volatilities.  For instance, \citet{billio2012econometric} constructs financial networks by assessing bivariate Granger causal relationships between firms' monthly stock returns.  \citet{basu2017system} refines these methods by employing \textit{multivariate} Granger causality, which has the effect of removing indirect edges from the network.  Similarly,  \citet{karpman2022exploring} uses multivariate \textit{quantile} Granger causality to estimate inter-firm connections that occur specifically during market downturns.  \citet{diebold2014network}, on the other hand, models daily return volatilities using VAR models, with the corresponding forecast error variance decompositions defining edges in the network. 

Most of the literature on financial network analysis has used low-frequency data such as monthly, weekly, or daily returns and volatilities.  However, firm linkages that are estimated from -- for example -- monthly financial data are challenging to interpret since it is difficult to establish what mechanisms, over the course of the month, produce those linkages (investment decisions are usually made within a much shorter time frame). Analyzing high-frequency, intraday financial data has the potential to yield insights that we cannot obtain through a low-frequency lens.  With the rise of high-frequency computerized trading, financial data is now being recorded at the level of nanoseconds, yielding massive \textit{intraday} datasets.  For example, the New York Stock Exchange (NYSE) maintains the Trade and Quote (TAQ) database, which provides detailed information (e.g., timestamp, price, size, etc.) on all trades and quotes for stocks that are active on U.S.-based exchanges.  By considering high-frequency financial data, we can better discern which aspects of a firm's trading give rise to its associations with other firms. One notable exception to the preponderance of low-frequency analyses is the work of \citet{hardle2018time}, which uses intraday limit order book data (bid and ask prices and volumes) to measure stock connectedness.

While using intraday data provides interpretability benefits, high-frequency financial time series also pose significant modeling challenges that are not present at lower frequencies [\citet{dutta2022review}].  These time series are often nonlinear and nonstationary, exhibiting strong persistence, seasonal and intraday patterns, and volatility bursts.  As a result, simple linear models cannot capture key features of the data.  Moreover, increasingly sophisticated financial products and trading algorithms have rendered the markets so complex that specifying a functional form (such as a linear model) to relate firms' variables may be overly simplistic.  

In this work, we adopt a nonlinear, nonparametric approach to estimate predictability across firms' time series using high-frequency data. Our methodology does not impose a functional form on the dynamic relationships between firms, and offers greater modeling flexibility. In addition, we move away from price-based measures (e.g. stock returns and volatility) and use  trade-based measures, which are expected to contain more fine-grained information.

In particular, we expand on the methodology proposed in \cite{microMachine}, which uses a random forest to predict a set of \textit{market measures}, variables that traders use as inputs to their execution algorithms, including measures of liquidity, volatility, and the shape of the returns distributions.  The features of their (and our) random forest are \textit{microstructure variables}, quantities that are computed from the price and volume of trades and that reflect underlying market frictions.  We note that most analyses consider only price data; for instance, the stock returns used in \citet{billio2012econometric} and realized volatility used in \citet{diebold2014network} can be calculated from stock prices.  However, trades are the truly fundamental object since they are what give rise to prices.  Thus, by using microstructure measures, which are formed from trade information, we address whether there is information contained in trades -- beyond what is reflected in prices -- that is useful in understanding inter-firm connections.

Easley et al. (2019) provides empirical evidence that random forests can predict market measures of futures contracts using those contracts' microstructure variables.  Their analysis focuses mainly on intra-firm prediction; that is, they consider whether contract A's microstructure variables can predict contract A's market measures, and similarly for contracts B, C, etc.  In our work, on the other hand, we ask whether features of firm A can help predict the market measures of firm B, for all pairs, A and B, in the system.  

Put differently, we measure whether, and to what extent, firm A's predictability increases when we include firm B's features in the random forest, as compared to when only firm A's features are used. Various metrics exist to quantify predictability, including accuracy, precision, recall, and the F1 score.  We use the area under the ROC curve (AUC), which reflects the true and false positive rates as the decision threshold is varied.  We apply a bootstrap procedure to test by how much (if at all) the AUC increases when we add firm B's features. The increase is then used as a weight for the edge running from firm B to firm A. In this manner, we construct a network whose edges indicate cross-predictability between firms.  This technique can be viewed as a high-frequency analogue to the Granger causality methods applied to monthly stock returns in \citet{billio2012econometric}. Under that framework, an edge from firm B to firm A means that firm B's lagged returns help predict firm A's returns, over and above firm A's own lagged returns. Edges are defined similarly here, except that instead of using linear models on monthly stock returns, we apply random forest methods to intraday data.  In both cases, we assess whether another firm's information boosts predictive power.  We note that random forests may be a particularly effective method for capturing cross-effects between firms since they allow for higher-order interactions between many features, which is especially important given the complexities of modern-day financial markets.

We apply our methodology to high-frequency trade data of U.S. banks, broker-dealers, and insurance companies, with the goal of better understanding cross-effects between these institutions.  Our methods can be used to address the same questions that researchers ask in the low-frequency context, including how network connectivity changes over time and what information channels exist between firms.  On the first count, we apply our methods to intraday data spanning 1998 to 2010, thereby visualizing the historical evolution of network connectivity over both economically stable and crisis periods. We find that the networks reach maximum density in late 2007, following the collapse of two subprime mortgage funds associated with the investment bank Bear Stearns.  Several of the most highly connected nodes in the network, including Lehman Brothers and AIG, have been recognized as key contributors to the U.S. financial crisis. Second, we demonstrate how our methods can be used to detect possible information spillovers between small and large financial firms.  This line of analysis is motivated by \cite{chordia2011liquidity}, which provides empirical evidence that the returns of large stocks lead (in the Granger causal sense) the returns of small stocks, and that this lead-lag relationship is especially strong when the large stocks have low liquidity. Our results are consistent with this earlier analysis: we find that the microstructure variables of large firms tend to be more important (compared to the microstructure variables of small firms) in predicting market measures for both small and large firms.

The manuscript is organized as follows.  In Section \ref{sec:ML_HFT_methods}, we describe our methodology, including an overview of how the data is structured, an explanation of how random forests work, and details of our bootstrap AUC procedure, which is used to assess whether cross-features provide predictive improvement.  In Section \ref{sec:ML_HFT_market_micro_variables}, we define five microstructure variables that are used as features in the random forest, while Section \ref{sec:ML_HFT_market_measures} presents two market measures that serve as labels (variables that we predict).  In Section \ref{sec:ML_HFT_data}, we describe the high-frequency data used in our analysis, including which firms we choose to focus on.  Section \ref{sec:ML_HFT_results} presents the results of two empirical analyses, namely the evolution of network connectivity over time and the presence of information flows between small and large firms.  Section \ref{sec:ML_HFT_discussion} concludes.

\section{Methods}
\label{sec:ML_HFT_methods}

In this section, we first describe how our data is structured, namely how we sample from high-frequency trade information to construct microstructure variables and market measures.  We then outline our random forest methodology, including an overview of how random forests work and details of our training and testing procedure.  Lastly, we introduce two metrics used to interpret the random forest results; the first measures the relative importance of features used in the model, while the second quantifies the random forest's predictive accuracy.  By comparing the accuracy with and without cross-features, we create financial networks whose edges indicate that one firm's variables significantly improve our ability to predict the other firm's market measure.  Details are provided below.

\subsection{Data Structure}
\label{sec:ML_data_structure}
Before describing our statistical methods, we briefly explain how our dataset is structured.  We begin by obtaining high-frequency trade data from the NYSE Trade and Quote (TAQ) database [\citet{TAQ}].  TAQ provides information on every trade that occurs on a U.S.-based exchange, including the NYSE, Nasdaq, National Stock Exchange, and others.  Among the many variables returned by TAQ are the \textit{timestamp}, \textit{price}, and \textit{volume} of each trade.  These three variables are integral to creating our final dataset: \textit{timestamps} are used to aggregate trades (thereby reducing the total number of observations in our dataset), while \textit{price} and \textit{volume} are used to create the microstructure variables and market measures that serve as features and labels in our random forest.

\subsubsection{Trade Aggregation}
\label{section: tradeAgg}
Aggregating trades is common in high-frequency financial data analysis, for several reasons: aggregation limits the effect of noise, reduces the amount of data that we need to process, and allows for the creation of economically meaningful variables [\citet{hautsch}]. Trade aggregation can be based on time (e.g., collecting all trades whose timestamps fall in a 30-minute interval) or on event (e.g., aggregating trades until the price change exceeds a given threshold).  In our analysis, we use time-based aggregation, grouping each firm's trades into 30-minute \textit{time bars}.\footnote{One alternate sampling method is to collect trades until their cumulative dollar-volume reaches a certain level [\citet{microMachine}].  So-called \textit{dollar-volume bars} have appealing theoretical and practical properties; however, they are not synchronized across stocks and thus present challenges for how to model using cross-effects.  For example, an actively traded stock, $s_A$, fills its dollar-volume bars faster than a less actively traded stock, $s_I$.  Thus $s_A$'s first dollar-volume bar may run from 9:30 AM to 9:35 AM, while $s_I$ does not fill its bar until 10:00 AM.  Therefore, we cannot use dollar-volume bars if we hope to use $s_I$'s features to make predictions about $s_A$: in effect, we would be using future information about $s_I$ to predict current properties of $s_A$.}   Since we consider only trades that occur during regular market hours (9:30 AM EST to 4:00 PM EST), our time bars correspond to the intervals 9:30 AM to 10:00 AM, 10:00 AM to 10:30 AM, and so on and so forth until 3:30 PM to 4:00 PM, with these bars repeated for each day of the sample period.  We emphasize that time bars are formed on a firm-by-firm basis; that is, we do not combine trades of stocks $x$ and $y$ into a single bar.

\subsubsection{Microstructure and Market Variables, Lookback Windows, and Forecast Horizons}
Once a firm's trades have been gathered into time bars, we construct a set of microstructure variables and market measures that capture key properties of the firm's trading. In Sections \ref{sec:ML_HFT_market_micro_variables} and \ref{sec:ML_HFT_market_measures}, we provide definitions of these variables and measures.  For now, we note that microstructure variables are used as \textit{features} (predictors) in our random forest, while market measures are used to calculate \textit{labels} (quantities we predict).  All are based on sequences of trade prices and volumes, and all are computed -- at each time bar -- using a lookback window of size $W$.  For instance, the value of Kyle's lambda (one of the microstructure variables) at time bar $t$ is based on the trade prices and volumes at time bars in $\{t, t-1, ..., t-W+1\}$.  

The microstructure variables at time bar $t$ are then used to predict the \textit{sign of the change} in a market measure at time bar $t+h$, where $h$ is a fixed forecast horizon.  For example, one of the market measures we consider is realized volatility.  We do not predict the value of realized volatility at bar $t+h$, nor do we predict the change in realized volatility between bars $t$ and $t+h$.  Instead we predict whether this change is positive (realized volatility increases) or negative (realized volatility decreases).  The sign of the change in realized volatility becomes the label for our random forest; thus we are predicting a binary variable that takes the value 1 if the market measure increases and -1 if it decreases.

In our analysis, we set $W = 50$ and $h = 50$.  Since each time bar represents a 30-minute interval and there are 12 such intervals during regular market hours\footnote{The market is open from 9:30 AM EST to 4:00 PM EST, which corresponds to 13 30-minute intervals.  However, we remove the first time bar of the day (see Section \ref{sec:ML_HFT_data} for details), resulting in only 12 time bars per trading day.}, our lookback window size and forecast horizon both correspond to slightly more than four trading days.

\subsection{Random Forest}
Random forests are a popular machine learning tool for predicting the values of a binary variable [\citet{breiman2001random}], 
 \citet{friedman2001elements}]. In our work, this binary variable represents whether a market measure — such as realized volatility — decreases (-1) or increases (1) over some fixed forecast horizon.  Random forests work by aggregating the predictions of many \textit{decision trees}, so we begin by describing how each tree makes its prediction.

A decision tree takes as input training data in the form $\{(x_i, y_i)\}_{1 \leq i \leq n}$, where $y_i$ is the label for observation $i$ and $x_i = (x_{i1},...,x_{ip})$ is the vector of features. The tree repeatedly splits the $n$ training observations into two subsets on the basis of one of the $p$ features.  For example, the first split might separate the training set based on whether the second feature is greater than 5.  This would yield two subsets, $\{j: x_{j2} \leq 5\}$ and $\{j: x_{j2} > 5\}$.  The next split could be based on whether the third feature is greater than 10, yielding 4 subsets, 
\begin{align*}
&\{j: x_{j2} \leq 5, x_{j3} \leq 10\} \hspace{20mm} \{j: x_{j2} \leq 5, x_{j3} > 10\}\\
&\{j: x_{j2} > 5, x_{j3} \leq 10\} \hspace{20mm} \{j: x_{j2} > 5, x_{j3} > 10\},
\end{align*}
and so on and so forth.  In this example, features 2 and 3 are referred to as \textit{split features}, while 5 and 10 are \textit{split points}.  Decision trees choose split features and split points by maximizing information gain, which measures how pure the labels are in the subsets that result from the split.  Maximum purity (information gain) is achieved when one subset contains only observations with label 1 and the other subset contains only observations with label -1. As we move further down the tree, generating more and more splits, the feature space becomes increasingly partitioned and there exist fewer observations in each node of the tree. Eventually the tree stops growing (according to a particular stopping criterion) and we classify each observation by considering the terminal node (aka leaf) to which that observation belongs. Specifically, each observation is predicted to have the most commonly occurring label (-1 or 1) in its leaf.

Decision trees are known to have low bias and high variance [\citet{friedman2001elements}].  They are accurate, on average, but individual decision trees are prone to overfitting the training data and sometimes do not perform well when generalized to a test set. Random forests counteract overfitting by aggregating the predictions of many decision trees, thereby stabilizing the overall prediction [\citet{breiman2001random}]. In particular, for each observation $i$, the random forest computes the fraction of trees that predict -1 vs. 1.  We can then make a prediction for observation $i$ based on which class has the higher probability.  For example, suppose a random forest consists of 100 trees, 55 of which predict -1 and 45 of which predict 1, yielding class probabilities of 0.55 and 0.45.  Then we can set our final prediction to be -1 since it is the majority vote over all trees.  Each decision tree in the forest is trained on a bootstrapped sample; that is, we draw $K$ samples with replacement from our training set and fit $K$ decision trees, one on each bootstrap sample. 

Lastly, an important aspect of random forests is that not all features are taken to be candidates for every split. Instead -- at each split -- we choose a random subset of features, compute the largest information gain we can achieve with each of these features (over all split points), and select as our final split feature and split point the ones that offer maximal information gain. This procedure is particularly helpful when there are correlated features, in which case decision trees may select the feature that offers marginally higher information gain, while ignoring its highly correlated but slightly less predictive counterpart. By randomizing the split candidates, we ensure that each of these feature has an equal opportunity of being selected.

\subsubsection{Random Forest Parameters}

We implement our random forest using the $\texttt{randomForest}$ package in $\texttt{R}$ [\citet{randomForestPackage}].  In particular, we produce forests with $K = 1000$ trees, each of which is allowed to grow without limit (i.e., the minimum leaf size is 1).  We randomly select $m$ features as candidates at each split, with $m = \left \lfloor{\sqrt{p}}\right \rfloor$.  Recall that $p$ denotes the number of features, which varies based on whether we include cross-effects.  We use five microstructure variables (see Section \ref{sec:ML_HFT_market_micro_variables}), so $m$ is either $\left \lfloor{\sqrt{5}}\right \rfloor$ if we use only the firm's own variables for prediction or $\left \lfloor{\sqrt{10}}\right \rfloor$ if we also consider the features of one other firm.  Finally, we assign a weight to each training set observation on the basis of its class; observations with label 1 (resp. -1) have weight $1/n_I$ (resp. $1/n_D$), where $n_I$ and $n_D$ denote the number of training observations with label 1 and -1, respectively.  When performing bootstrap, training observations are randomly sampled according to these weights so that the effect of class imbalance is minimized.

\subsubsection{Purged Cross-Validation}

\begin{figure}[ht]
    \centering
    \includegraphics[trim = {6cm 10.5cm 6cm 3cm},clip]{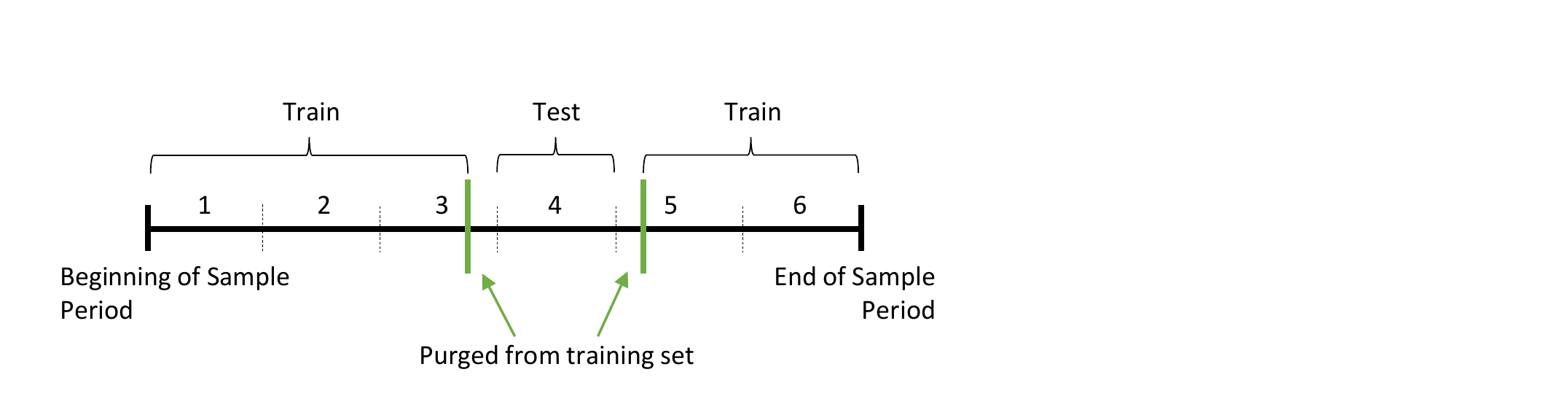}
    \caption{Schematic of the purged cross-validation procedure.  The sample period is divided into 6 intervals of equal length, each interval serving as a test set as we iterate over the sample period.  Suppose interval 4 is the current test set.  Then five days worth of data is purged from before and after interval 4, and the remaining data is used as the training set.}
    \label{fig:purged_CV}
\end{figure}

Once the random forest is fit on the training set, we evaluate its performance on test data.  Depending on the exact analysis we perform (see Section \ref{sec:ML_HFT_results} for details), we use one of two approaches.  The first procedure is purged cross-validation as proposed in \citet{microMachine}.  This involves splitting the sample period into $G$ intervals of equal length.  We then iterate over each interval, $g$, taking $g$ to be the test set and using all other intervals as training data, with one caveat.  Since our microstructure variables (features) and market measures (labels) are formed using a lookback window, the train and test sets under this approach are not independent of each other, introducing bias into our results.  To correct for this, we purge five days worth of data from around  each test set, $g$ (see Figure \ref{fig:purged_CV}).  This procedure yields $G$ sets of results, one for each interval.  In Section \ref{sec:ML_HFT_metrics}, we discuss how to aggregate these results across test sets.

Purged cross-validation allows us to test on the entire dataset as we iterate over intervals; however, it has the disadvantage that sometimes we test our model on data that occurs \textit{prior to} our training data.  (This would not be the approach of, say, a practitioner applying a random forest to recent financial data in order to forecast changes in market measures.)  To ensure that the chronology of the train and test sets does not impact our final results, we use an alternative approach for some of our analyses.  This consists of splitting the sample period into two intervals, training on the earlier interval (which has some data purged from it) and testing on the later interval.

\subsection{Evaluating the Random Forest}
\label{sec:ML_HFT_metrics}

After applying our random forest to the test sets, we consider two aspects of our model's performance: (i) its predictive ability, i.e., how well the random forest classifies observations in the test set, and (ii) which features are most important in making those predictions.  We address each of these points in turn.  

\subsubsection{AUC for Assessing Prediction Accuracy and Forming Networks}
The receiver operating characteristics (ROC) curve offers a visual medium by which we can assess the predictive performance of a binary classifier such as a random forest [\citet{fawcett2006introduction}].  For each observation in the test set, the random forest provides the probability that the observation's label is -1, from which we can readily compute the probability that the observation's label is 1.  We convert these probabilities to actual predictions of -1 or 1 by setting a decision threshold and evaluating whether the observation's probability (e.g., of being -1) meets this threshold.  For instance, if we set the decision threshold to 0.5, then observations are classified according to whether the majority of trees in the random forest predict -1 or 1 for the observation in question.

The ROC curve displays the tradeoff between the true positive rate and the false positive rate as we vary the decision threshold between 0 and 1.  The true positive rate (also referred to as recall) is defined as
\begin{align}
\label{eq:TPR}
    TPR = \frac{TP}{TP + FN},
\end{align}
where $TP$ (resp., $FN$) is the number of true positives (resp., false negatives) produced by the classifier at a set threshold.  In our analysis, we take labels of 1 to be positives and -1, negatives.  From equation (\ref{eq:TPR}), we can see that the true positive rate is simply the proportion of positives in our system that are correctly classified as such.  Similarly, the false positive rate is given by
\begin{align}
\label{eq: FPR}
    FPR = \frac{FP}{FP + TN},
\end{align}
where $FP$ (resp., $TN$) is the number of false positives (resp., true negatives) produced by the classifier at a set threshold.  The false positive rate is then the proportion of negatives in our system that are incorrectly classified as positives.  Both the TPR and FPR can be computed given the predicted class probabilities and true labels for the test set.  Recall that, for purged cross-validation, we use multiple test sets; however, we simply aggregate the predicted and true values across all intervals, yielding -- in effect -- a single set of test results.

\begin{figure}[ht]
    \centering
    \includegraphics[angle = -90, scale = 0.5]{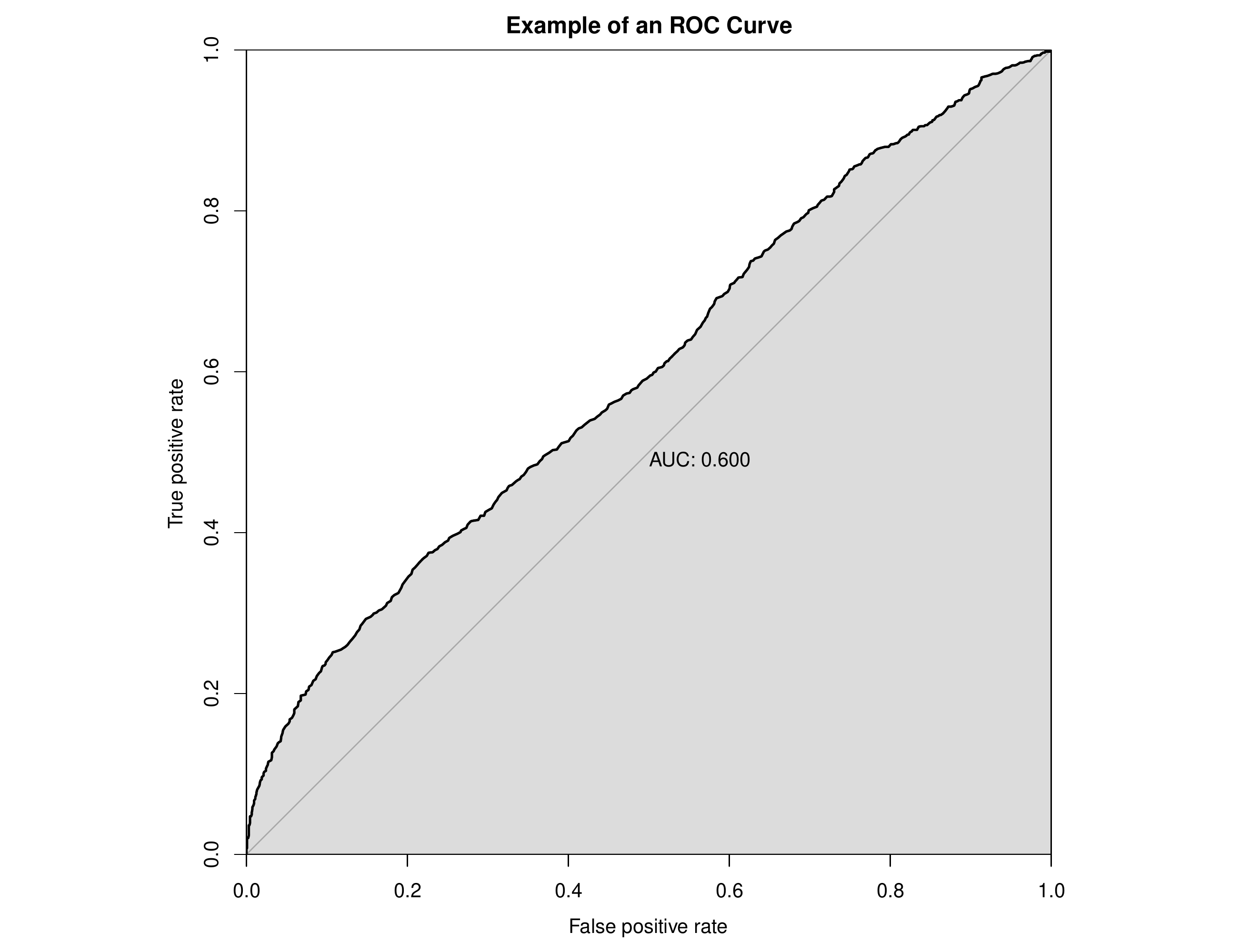}
    \caption{Receiver operating characteristics (ROC) curves plot the true and false positive rates of a binary classifier as the decision threshold is varied.  Random classifiers have a diagonal ROC curve, with a corresponding area under the curve (AUC) of 0.5.  Higher values of AUC, as illustrated here, indicate better classifier performance.}
    \label{fig:roc_curve}
\end{figure}

As we vary the decision threshold from 0 (all observations classified as positive) to 1 (all observations classified as negative), both the TPR and the FPR decrease from 1 to 0.  The ROC curve plots the TPR and FPR at each of these intervening thresholds (see Figure \ref{fig:roc_curve}).  A random classifier (i.e., one which -- for each observation -- predicts -1 or 1 with equal probability) yields a diagonal ROC curve running from $(0,0)$ to $(1,1)$, while a classifier that perfectly separates negatives from positives has an ROC curve running from $(0,0)$ up to $(0,1)$ and across to $(1,1)$.  Thus, we can quantify a classifier's performance by computing the area under the ROC curve, referred to as the AUC.  In the case of a random classifier, the AUC is 0.5, while a perfect classifier has an AUC of 1.  AUC has the advantage that it assesses a classification model's performance over all possible decision thresholds, without requiring us to set a single threshold.

We use AUC to detect the presence of cross-effects between firms; that is, to assess whether microstructure variables of firm $x$ are useful in predicting market measures of firm $y$.  We take the view that there are two competing models: \textit{Model 1} is a random forest not containing any cross-features (firm $y$'s variables only), while \textit{Model 2} is a random forest that does contain cross-features (both firm $x$ and firm $y$'s variables).  If features from firm $y$ have predictive power, then \textit{Model 2} should have a higher AUC than \textit{Model 1}.  Thus, to determine whether cross-effects exist, we test the following hypotheses:
\begin{align}
\label{eq:auc_test}
    H_0: AUC_2 = AUC_1 \hspace{10mm} \text{vs.} \hspace{10mm} H_A: AUC_2 > AUC_1,
\end{align}
where $AUC_\ell$ denotes the AUC of Model $\ell$, with $\ell = 1,2$.\footnote{Note that we consider only whether $AUC_2$ is greater than $AUC_1$ since the reverse does not indicate the presence of cross-effects.  In theory, the AUC should either (a) increase if we include microstructure measures from a firm having predictive power, or (b) stay the same if we include microstructure measures from a firm that does \textit{not} have predictive power (i.e., the random forest should be able to select -- as split features -- the microstructure measures that improve predictive performance, so that the AUC remains the same with the addition of an ``unhelpful'' firm, but does not decrease).}  
The test in (\ref{eq:auc_test}) is executed according to the following steps:
\begin{enumerate}
     \item We fit \textit{Models 1} and \textit{2} on the training data, apply the fitted models to the test data, and store $\left\{P^1_i\right\}_i$, the predicted class probabilities from \textit{Model 1}; $\left\{P^2_i\right\}_i$, the predicted class probabilities from \textit{Model 2}; and the true test set labels, $\left\{R_i\right\}_i$.  Here $i$ indexes observations in the test set.  
     \item Using the predictions and true values, we compute $AUC^O_1$ and $AUC^O_2$, the areas under the curve for \textit{Models 1} and \textit{2}, respectively. 
     \item We then draw $B$ bootstrap samples from $\left\{P^1_i, P^2_i, R_i\right\}_i$.  For each bootstrap sample $b$, with $b = 1,...,B$, we calculate new areas under the curve, $AUC^b_1$ and $AUC^b_2$, storing the difference, $AUC^b_2 - AUC^b_1$.
     \item The standard deviation, $s$, of the bootstrap differences is computed and a test statistic, $D$, is calculated as
\begin{align*}
    D = \frac{AUC^O_2 - AUC^O_1}{s}.
\end{align*}
\item Finally, a one-sided $p$-value is computed under the assumption that $D$ follows a normal distribution.\footnote{This bootstrap procedure is implemented using the \texttt{roc.test()} function in the \texttt{pROC} package within \texttt{R} [\citet{pROC}].  We set $B = 2000$.}
\end{enumerate}
Steps 1-5 are repeated twice for each pair of firms, $(x,y)$, in the system, once to make predictions for firm $x$ and again to make predictions for firm $y$.  This yields a set of $2 \times {N \choose 2}$ $p$-values, where $N$ is the number of firms under consideration.  We apply a multiple testing correction to control the false discovery rate [\citet{benjamini1995controlling}] and form directed networks with edges between pairs of firms that have an adjusted $p$-value falling below some threshold.

\subsubsection{MDA for Feature Importances}

Area under the curve measures the random forest's predictive performance; however, we are also interested in knowing to what extent the various features contribute to these predictions.  We quantify feature importances using the \textit{mean decrease in accuracy} (MDA), which compares the random forest's accuracy on the original data to its accuracy on a dataset for which the values of a feature have been randomly permuted [\citet{biau2016random}].  Accuracy is defined as the fraction of all test set observations that are classified correctly\footnote{Here we use a decision threshold of 0.5 to convert probabilities to predicted values.}:
    \begin{align*}
        A = \frac{TP + TN}{TP + FP + TN + FN}.
    \end{align*}
For each feature $f$, we compute its MDA as follows:
\begin{enumerate}
    \item We begin by fitting a model to the training set and computing its accuracy, $A_O$, on the test set. 
    \item Next, we randomly permute the values of $f$ in the test set.  We make predictions on this shuffled test set and compute the new accuracy, $A_P$.
    \item The MDA for feature $f$ is the fraction by which the model's test set accuracy decreases after shuffling $f$:
    \begin{align}
    \label{eq:MDA}
        MDA = \frac{A_O - A_P}{A_O}.
    \end{align}
\end{enumerate}
Features having a high MDA are considered to be more important since they have a large effect on the model's accuracy.  In our analysis, we compute the MDAs separately for each test set in the sample period.  The MDA values are then averaged over test sets to yield a mean importance for each feature $f$.

\section{Market Microstructure Variables}
\label{sec:ML_HFT_market_micro_variables}

Our random forest model uses a variety of market microstructure variables as features.  Microstructure variables are designed to measure illiquidity, volatility, order imbalance, and other consequences of market frictions.  As in \citet{microMachine}, we focus on five such measures that represent the evolution of microstructure models from those that use price data alone (first generation) to those that use both price and volume data (second generation) to those that use more extensive trade information (third generation).  Most of these measures were designed before the advent of high-frequency trading, raising the question of how well they capture market frictions in our current, more complex financial era.  Thus our model helps to assess the ongoing utility of these traditional market microstructure variables.  In what follows, we describe each of the five measures, including their importance and how they are computed. 

\subsection{Roll Measure}
The Roll measure -- a first generation microstructure variable -- uses sequences of price changes to estimate the effective bid-ask spread, which in turn is a proxy for the transaction cost [\citet{hautsch}].  The Roll measure at bar $t$, written $R_t$, is a function of the first-order serial covariance of price changes:
\begin{align}
    R_t = 2\sqrt{\big\vert cov(\boldsymbol{\Delta p_t}, \boldsymbol{\Delta p_{t-1}}) \big\vert}.
\end{align}
Here $\boldsymbol{\Delta p_t} = \left(\Delta p_{t-W}, \Delta p_{t-W+1}, ..., \Delta p_{t-1}, \Delta p_t\right)$, with $\Delta p_t$ denoting the difference between the closing prices at bars $t$ and $t-1$.

\subsection{Roll Impact}
Roll impact, a second generation variable, is closely related to the Roll measure.  Specifically, Roll impact is defined as the Roll measure, scaled by the amount of dollar-volume traded over the bar:
\begin{align}
    \tilde{R}_t = \frac{R_t}{\sum_{k \in \mathcal{T}(t)} p_kv_k},
\end{align}
where $\mathcal{T}(t)$ is the set of trades belonging to bar $t$, and $p_k$ and $v_k$ are the price and volume, respectively, of trade $k$.  Since the numerator, $R_t$, represents transaction cost, Roll impact can be interpreted as the \textit{transaction cost per unit of trade}.

\subsection{Kyle's Lambda}
Kyle's lambda at bar $t$ is given by
\begin{align}
    \lambda_t = \frac{p_t - p_{t-W}}{\sum_{\tau = t-W}^t b_\tau V_\tau},
\end{align}
where $p_t$ is the closing price of bar $t$, $V_t$ is the total volume traded over bar $t$, and $b_t = sign(p_t - p_{t-1})$.  Kyle's lambda is the coefficient obtained by regressing price change on order flow, and thus measures the price impact of trading.

\subsection{Amihud's Lambda}
Amihud's lambda, another second generation variable, measures illiquidity by computing the ratio of the price change to the amount traded.  Thus Amihud's lambda can be viewed as the ``price change per trade size,'' with less liquid assets having a larger per-unit price impact than their more liquid counterparts [\citet{hautsch}]. In particular, Amihud's lambda at bar $t$ is defined as
\begin{align}
    \lambda^A_t = \frac{1}{W}\sum_{\tau = t - W + 1}^t \frac{|r_\tau|}{\sum_{k \in \mathcal{T}(\tau)} p_kv_k},
\end{align}
where $r_t$ is the return over bar $t$.

\subsection{VPIN}
The volume-synchronized probability of informed trading (VPIN) arises from third generation market microstructure models.  By comparing the amount of buyer- and seller-initiated trades, VPIN quantifies the extent to which there is information asymmetry in the market.  For example, if a group of traders knows that an asset's price is about to rise, we may observe a preponderance of buyer-initiated trades as informed traders rush to secure the asset before its price increases.  The VPIN at bar $t$ is given by 
\begin{align}
    VPIN_t = \frac{1}{W}\sum_{\tau = t - W + 1}^t \frac{\big\vert \hat{V}^S_\tau - \hat{V}^B_\tau\big\vert}{V_\tau},
\end{align}
where $V_t$ is the total volume traded over bar $t$, $\hat{V}^B_t$ is the estimated total buy volume over bar $t$, and $\hat{V}^S_t = V_t - V^B_t$ is the estimated total sell volume over bar $t$.  Importantly, the information provided in the Trade and Quote (TAQ) database does not include whether the trades were buyer- or seller-initiated (we call such trades ``unsigned'').  Thus, before computing the VPIN, we must first classify trades as buys or sells.  A number of methods exist for this purpose (e.g, the Lee-Ready algorithm and the tick rule); here we use bulk volume classification (BVC), which has been demonstrated to outperform other techniques when the trade data is noisy [\citet{BVC}].

\subsubsection{Bulk Volume Classification}
Bulk volume classification is based on the heuristic that, if a trade is buyer-initiated, it will take place at the ask (the lowest price offered by sellers) and therefore will generate an uptick in the price of the asset.  Similarly, if the trade is seller-initiated, it will take place at the bid (the highest price offered by buyers) and therefore will produce a downtick in price.  This idea suggests that we can determine the amount of buyer- (resp., seller-) initiated trades by considering whether the price of the asset goes up or down.  More specifically, let $V_t$ be the total volume traded over bar $t$, with $\Delta p_t = p_t - p_{t-1}$ denoting the change in the closing price between bars $t$ and $t-1$.  Then BVC estimates the volume of buyer-initiated trades over bar $t$ to be
\begin{align}
\label{eq: BVC}
    \hat{V}^B_t = V_t \cdot \Phi\left(\frac{\Delta p_t}{\sigma_{\Delta p_t}}\right),
\end{align}
where $\sigma_{\Delta p_t}$ is the empirical standard deviation of the price changes (over all bars) and $\Phi$ is the cumulative distribution function of a standard normal random variable.  Notice that the more \textit{positive} the scaled price change is, the closer $\Phi\left(\frac{\Delta p_t}{\sigma_{\Delta p_t}}\right)$ is to 1, so that most of the volume traded over bar $t$ is classified as buyer-initiated.  Similarly, the more \textit{negative} the scaled price change, the more volume is classified as seller-initiated.  This result comports with the heuristic we described above: buyer-initiated trades are more likely to produce positive price changes, while seller-initiated trades are more likely to generate negative price changes.
\section{Market Measures}
\label{sec:ML_HFT_market_measures}
We use the above-described microstructure variables as features in our random forest, with the aim of predicting several important \textit{market measures}.  Although there are a number of market measures that interest traders, regulators, and researchers, here we focus on two: the sign of the change in realized volatility and the sign of the change in the kurtosis of returns.  We describe each in turn, explaining why they are of interest and how we compute them.  

\subsection{Sign of the Change in Realized Volatility}
Realized volatility is measured by the empirical standard deviation of returns; that is, if $r_t$ denotes the return over bar $t$, then the realized volatility, $\sigma_t$, is given by $\sigma_t = sd\left(r_{t-W+1}, r_{t-W+2}, ..., r_{t-1}, r_t\right)$.  The sign of the change in realized volatility is defined as
\begin{align}
    sign(\sigma_{t+h} - \sigma_t),
\end{align}
which is 1 when the realized volatility increases (over a forecast horizon of $h$ bars) and -1 when the realized volatility decreases.  A trader who predicts that volatility will rise may want to adjust their execution algorithm, increasing their trading activity so that orders are completed before prices begin to fluctuate [\citet{microMachine}].

\subsection{Sign of the Change in the Kurtosis of Returns}
Many standard risk models assume normally distributed returns; thus, traders are interested in forecasting any deviations from normality so that they can adapt their risk management practices accordingly.  One such deviation could be an increase or decrease in the kurtosis (``tailedness'') of the returns.  For example, high forecasted kurtosis could be caused by a drop in liquidity: with fewer orders on the book, trades are executed at more extreme prices, thereby generating more extreme returns [\citet{microMachine}].  The (excess) kurtosis\footnote{The kurtosis of the normal distribution is 3, meaning that the \textit{excess} kurtosis measures the ``tailedness'' of a given distribution relative to the normal distribution.  The terms ``kurtosis'' and ``excess kurtosis'' are often used interchangeably; thus, we simply refer to ``kurtosis.''} at time $t$ is given by 
\begin{align}
    K_t = \frac{\mu_{t,4}}{\sigma_t^4} - 3,
\end{align}
where $\mu_{t,4}$ and $\sigma_t$ are, respectively, the empirical fourth moment and standard deviation of $\left(r_{t-W+1}, r_{t-W+2}, ..., r_{t-1}, r_t\right)$.  The sign of the change in kurtosis is then 
\begin{align}
    sign\left(K_{t+h}-K_t\right).
\end{align}
\section{Data Description}
\label{sec:ML_HFT_data}
We obtain intraday trade data from the NYSE Daily Trade and Quote (TAQ) database, via Wharton Research Data Services (WRDS) [\citet{TAQ}].  TAQ includes trade and quote information for all stocks that are actively traded on a U.S.-based exchange; however, we focus our attention on firms from the financial sector, specifically banks, primary broker-dealers, and insurance companies.  In so doing, we are able to compare our results to the analyses in  \citet{karpman2022exploring}, where lower-frequency data (monthly returns) are used to construct financial networks on the same set of firms.  As in  \citet{karpman2022exploring}, sectoral membership of firms is identified using the Standard Industrial Classification (SIC) code.  We analyze data for this set of firms over two time periods: 1998-2010, and 2018 (see Sections \ref{sec:ML_evolutionConnectivity} and \ref{sec:ML_smallLargeAnalysis}, respectively).   

Starting with the full set of trades for these firms, we apply the following filters to compile our final dataset: (i) remove any trades whose price or volume is negative since these records are clearly erroneous, (ii) exclude trades occurring outside of regular market hours (9:30 AM to 4:00 PM EST), (iii) only retain trades of common shares\footnote{This corresponds to selecting records for which the TAQ symbol suffix is blank.}, and (iv) remove trades that are corrected, changed, or marked as erroneous\footnote{This corresponds to selecting records for which the TAQ trade correction indicator is ``00.''}.  For each stock, we form time series of each of the microstructure variables and market measures by aggregating trades into 30-minute time bars (see Section \ref{section: tradeAgg}).  Lastly, since the market opening is run according to a different process, namely, an auction, we remove the first bar of each day from our final dataset.   
\section{Results}
\label{sec:ML_HFT_results}

Having discussed the methods by which we construct high-frequency financial networks, we now demonstrate how such networks can be used to gain insight into the structure of the financial system using historical trade data.  We consider two examples.  The first examines how inter-firm connections vary over the course of 1998 to 2010, with a special focus on whether connectivity changes in and around financial crises (see Section \ref{sec:ML_evolutionConnectivity}).  The second example explores why edges appear between certain pairs of firms, and -- in particular -- whether the sizes of the firms (measured via market capitalization) plays a role (see Section \ref{sec:ML_smallLargeAnalysis}).

\subsection{Historical Evolution of High-Frequency Financial Network Connectivity}
\label{sec:ML_evolutionConnectivity}

Connections between financial institutions create channels through which risk can spread; hence, firm interconnectedness is considered to be a major contributor to \textit{systemic risk}, defined as the risk of widespread failure of the financial system.  For example, if a highly connected firm fails (even if due to an idiosyncratic shock), it may trigger a cascade of other firm failures that could cause extensive damage to the wider system.  In the years since the 2007-2009 U.S. Financial Crisis, there has been increasing interest in measuring systemic risk and in identifying systemically important financial institutions (SIFIs).  

Since systemic risk is tied to firm interconnectedness, much of the recent literature has explored how to use financial data (e.g., balance sheet information, returns, volatilities) to learn networks of firms.  For instance, \citet{billio2012econometric} and \citet{basu2017system} construct networks whose edges correspond to intertemporal correlations (Granger causality) between firms' stock returns.  Under the market efficiency hypothesis, there should not exist such lead-lag relationships between the price changes of different firms; however, in practice, market frictions such as capital requirements, borrowing constraints, and transaction costs may indeed give rise to correlations.  As argued in \citet{billio2012econometric}, the more such correlations exist (and the larger these correlations are), the greater the chance of risk propagating from one firm to another (i.e., the more systemic risk there is).       

\citet{billio2012econometric} shows that there are an increasing number of Granger causal connections during the economically unstable periods of 1998-1999 and 2007-2008.  Likewise, \citet{basu2017system} (which refines the methods in \citet{billio2012econometric}) demonstrates that network connectivity spikes around several recent systemic events, including the 1998 Russian financial crisis and the 2008 collapse of the investment bank, Lehman Brothers.  \citet{karpman2022exploring} expands on these methods further by constructing networks via quantile Granger causality, which focuses on firm connections that exist specifically during market downturns.  Each of the aforementioned studies uses monthly stock returns for network building.

Thus far we are unaware of any studies that attempt to quantify systemic risk using high-frequency financial networks.  The methods proposed in this paper, however, are a natural vehicle for doing so.  We have described how microstructure variables, computed from intraday trade data, can be used to predict future changes in market measures such as realized volatility.  Since these microstructure measures reflect information-based trading, a firm, $y$, having microstructure measures that can help predict realized volatility of another firm, $x$, represents a possible source of risk to firm $x$.  Thus, by assessing whether features from one firm are useful in forecasting changes for another firm, we can construct a network whose edges represent a high-frequency analogue of returns-based Granger causality.  In this section, we create such networks for a set of firms and over a given time period that are comparable to those considered in \citet{billio2012econometric}, \citet{basu2017system}, and  \citet{karpman2022exploring}.  We begin with the details of our network construction process, and then compare our results to those obtained using bivariate Granger causality applied to monthly stock returns.   

\subsubsection{Methodology for Constructing 1998-2010 Financial Networks}

For each year between 1998 and 2010, we rank all actively-traded firms according to their average monthly market capitalization, which is computed using data from the Center for Research in Security Prices (CRSP) database, accessed via WRDS [\citet{CRSP}].  Using this ranking, for each year, we identify the top 25 banks, primary broker-dealers, and insurance companies, yielding a total of 75 firms.\footnote{Our choice of firms is similar to that made in \citet{karpman2022exploring}, which considers companies in the same three financial sectors, but selects sets of firms over 36-month rolling windows, rather than on an annual basis.}  Any firms having insufficient data are excluded from our analysis, resulting in some variation in the number of firms considered per year (ranging from 59 firms in 1998 to 75 firms in the later years of the sample period).\footnote{As discussed in Section \ref{sec:ML_data_structure}, we aggregate trades into 30-minute time bars in order to create economic variables of interest, reduce the impact of noise, and decrease the amount of data inputted to the random forest.  For the earlier years in the sample period, some firms have 30-minute windows in which few trades occurred.  Out of an abundance of caution, we choose to exclude these firms from our analysis.  Specifically we discard any firm for which 25\% or more of its time bars contain fewer than 5 trades.}  

Next we divide each year into three overlapping 6-month periods: January 1 through June 30, April 1 through September 30, and July 1 through December 31.  Thus our analysis involves 39 time windows (13 years, with 3 windows per year).  For each six-month window, we split the interval into two sets, training on the first three months and testing on the last three months.  For example, we train on data from approximately\footnote{These dates are only approximate since we purge data from around the test set (see Section \ref{sec:ML_HFT_metrics}).} January 1, 1998 through March 31, 1998 and test on data from approximately April 1, 1998 through June 30, 1998.  We implement this testing procedure, rather than purged cross-validation, so as not to introduce any bias that may result from training on data that occurs \textit{after} the test data.

In each window, we iterate over each pair of firms, $(x,y)$, twice, once to predict the sign of the change in realized volatility for firm $x$, and a second time for firm $y$.  We fit two random forest models, one that includes only features of the firm for which we are forecasting and the other that includes cross-features (i.e., features of both $x$ and $y$).  Then, as described in Section \ref{sec:ML_HFT_metrics}, we use bootstrap to assess whether the area under the curve (AUC) increases significantly under the inclusion of cross-features.  Our bootstrap procedure yields a set of p-values, one for each possible edge, $x \rightarrow y$, in the network.  We apply a false discovery rate correction and retain the set of edges whose adjusted p-value is less than or equal to 0.05.

\subsubsection{Estimated 1998-2010 Financial Networks}

\begin{figure}[h]
    \centering
    \includegraphics[scale=0.28,angle=-90]{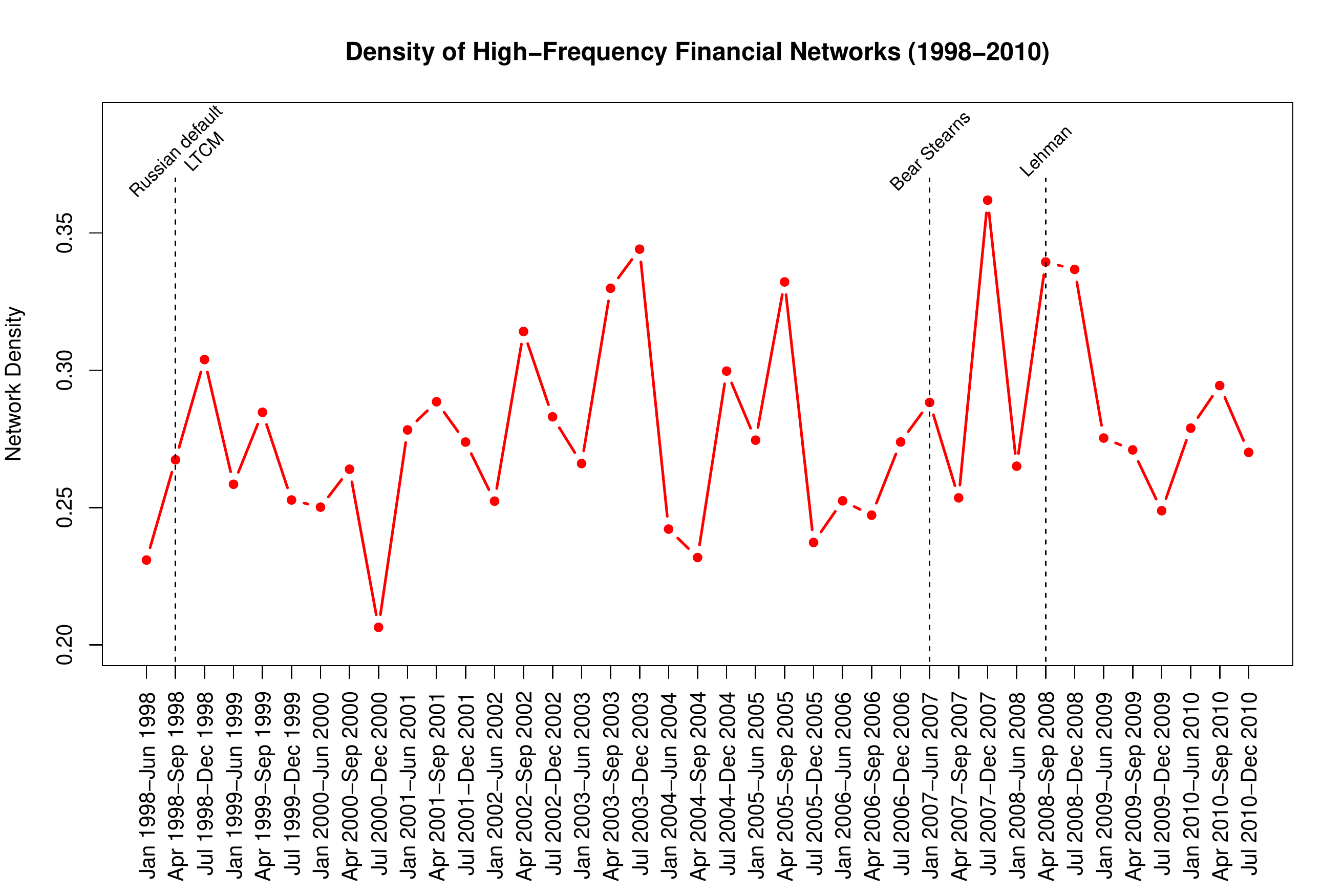}
    \includegraphics[scale=0.28,angle=-90]{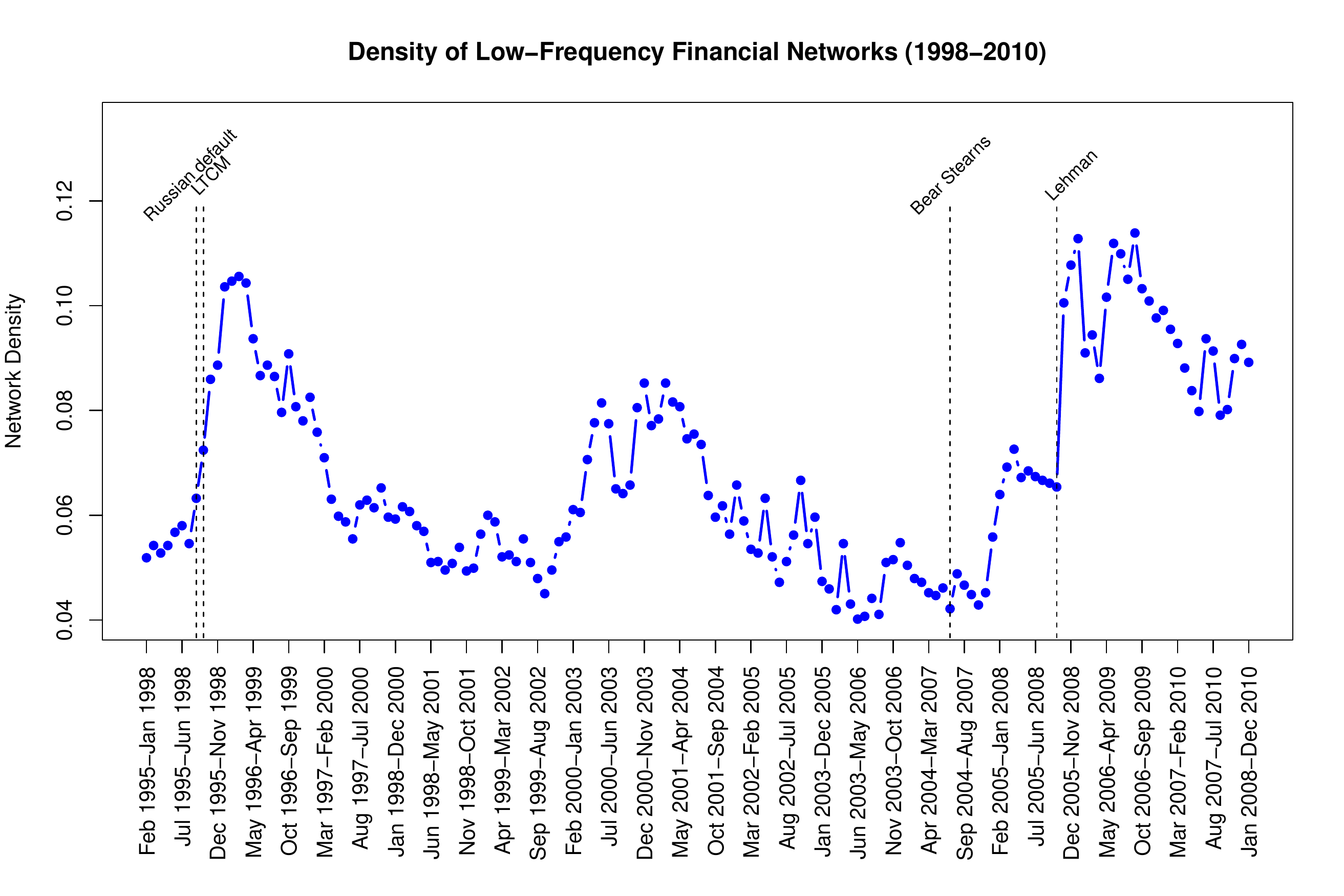}
    \caption{Density of financial networks over the 1998 to 2010 period, where network density refers to the proportion of realized edges.  Top (high-frequency networks): each year is divided into three overlapping windows of six months each, a network is estimated for each window by applying the methodology described in \ref{sec:ML_HFT_methods} to intraday trade data, and the network density is plotted.  Bottom (low-frequency networks): the sample period is divided into 36-month rolling windows, a network is estimated for each window by applying bivariate Granger causality to monthly stock returns, and the network density is plotted.  Note that there are 39 high-frequency networks and 156 low-frequency networks.}
    \label{fig:evolution_of_ntwk_connectivity}
\end{figure}

Figure \ref{fig:evolution_of_ntwk_connectivity} displays the proportion of realized edges, hereafter referred to as \textit{density},\footnote{For example, if half of all possible edges occur, then the density is 0.5.} in each of our estimated networks from 1998 to 2010.  For comparison purposes, we also show the density of networks estimated using bivariate Granger causality on monthly stock returns; however, we caution the reader that the high- and low-frequency networks are computed over different time windows, hence the two time series of network density are of different lengths.  

Our first observation is that the high-frequency network density increases steadily during 1998, reaching a peak in the last quarter of that year (i.e., when our model is applied to test data from October-December 1998).  This increase in connectivity coincides with a period of mounting economic turmoil in Russia, culminating with the Russian government devaluing the ruble, defaulting on domestic debt, and declaring a moratorium on repayment of foreign debt (August 17, 1998) [\citet{chiodo2002case}].  As the future of the Russian economy remained unclear, U.S. stocks plunged and the Federal Reserve Bank of New York was forced to organize a bailout of the U.S.-based hedge fund Long Term Capital Management [\citet{rubin1999hedge}].  Notice that low-frequency (monthly returns) networks also display a connectivity increase during the fall of 1998.  

Our high-frequency networks then become less dense through the end of 2000, at which point connectivity repeatedly increases and decreases (albeit with an overall upward trend) through late 2003.  These results are less interpretable than those in the low-frequency setting, where the density consistently decreases from 1999 through late 2002.  Both the low- and high-frequency networks have elevated density in 2003.  The intraday networks become less dense in 2004, before increasing in density in 2005.  On the other hand, the monthly-scale networks remain dense throughout 2003 and 2004 and are \textit{not} particularly dense in 2005.   

Intraday networks exhibit a fairly persistent increase in density through 2006 and 2007.  In fact, a global maximum density of 36.2\% is reached in the end of 2007, subsequent to the summer 2007 failure of two subprime mortgage funds associated with the investment bank Bear Stearns.  Connectivity then drops sharply in the first half of 2008 before increasing again.  These results are somewhat consistent with what is observed in the monthly-scale networks, where density steadily increases until the beginning of 2008, then decreases, and finally spikes following the September 2008 collapse of the investment bank Lehman Brothers.  High-frequency networks, like their low-frequency counterparts, display an overall decline in density in the late 2000s.

We now turn our attention to which firms are central in and around the 2007-2009 U.S. Financial Crisis.  Node centrality can be measured using a variety of metrics (e.g., degree, closeness, betweeness).  We focus on degree; that is, on how many edges are incident to the node. Firms can be characterized by both their in-degree (number of incoming edges) and out-degree (number of outgoing edges).  A firm having a large in-degree is one for which many other firms' microstructure measures are useful in forecasting its realized volatility.  On the other hand, a firm with a large out-degree has microstructure measures that are useful for predicting the realized volatility of many other firms.  Firms with large out-degree have the potential to spread risk through the financial system since aspects of their trading (captured via microstructure measures) propagate to other firms.  Likewise firms with large in-degree have the potential to absorb this risk.

\begin{figure}[!t]
    \centering
    \includegraphics[scale=0.35]{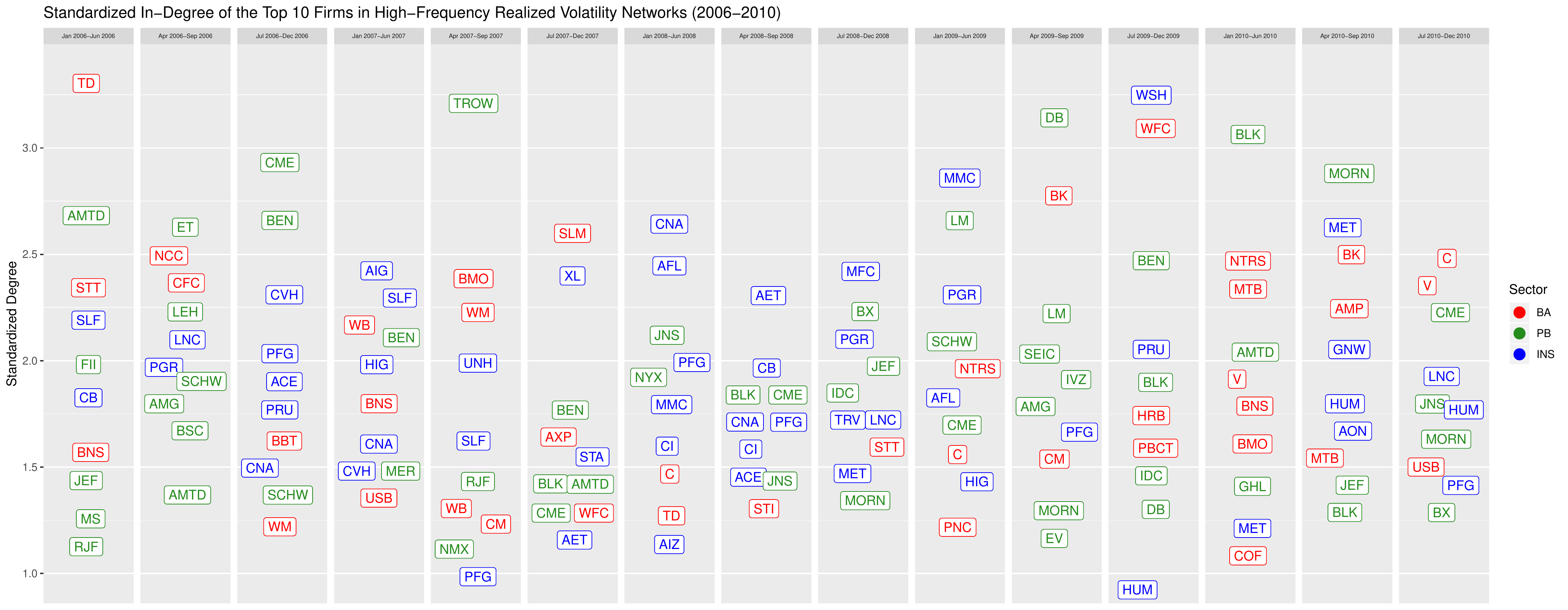}
    \includegraphics[scale=0.35]{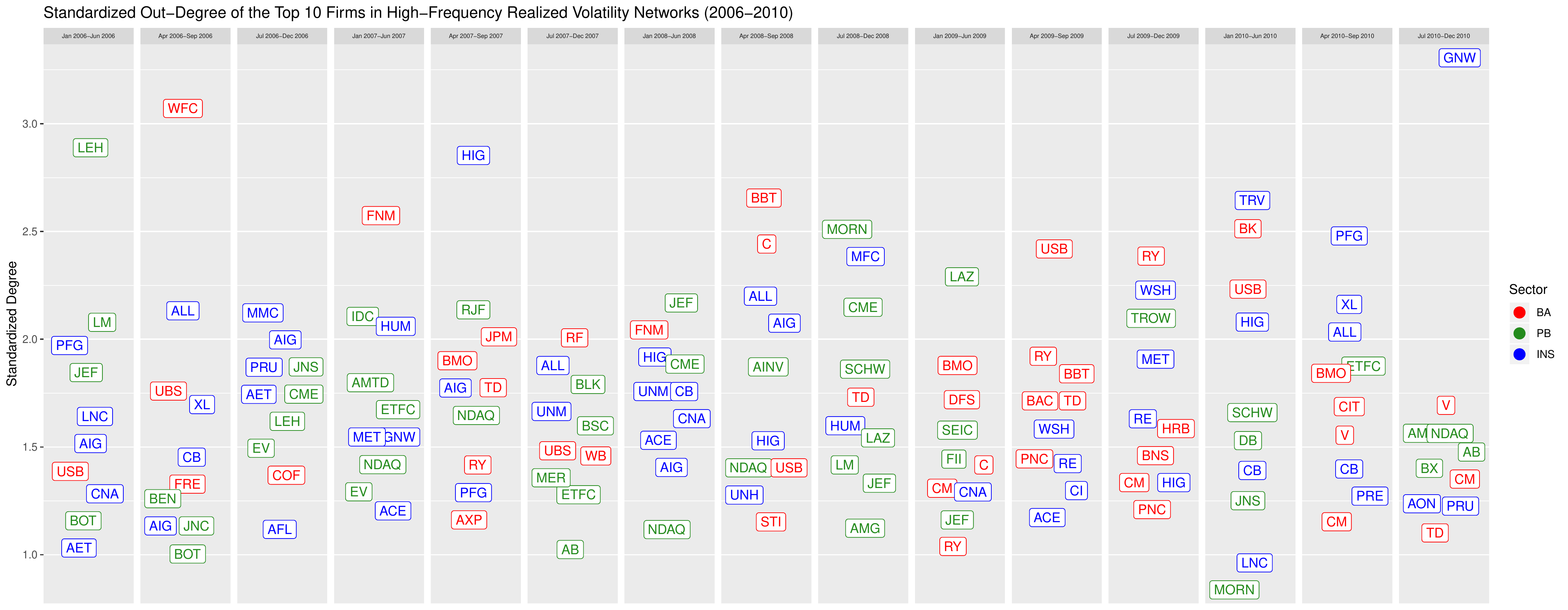}
    \caption{Highly connected firms in high-frequency realized volatility networks before, during, and after the U.S. Financial Crisis.  Firms are ranked according to their in-degree (top) and out-degree (bottom).  Standardized degrees (i.e., (firm degree - mean network degree)/standard deviation of network degree) are plotted so as to make results comparable across different networks.  Banks (resp., broker-dealers, insurance companies) are displayed in red (resp., green, blue).  Note that we add a small amount of random noise to the $(x,y)$ coordinates of each firm so that firm labels do not overlap with one another.  Full company names for all ticker symbols are provided in Table \ref{table: ML_HFT ticker symbol table}.}
    \label{fig:top_firms_high_freq}
\end{figure}

In Figure \ref{fig:top_firms_high_freq}, we display the 10 most highly connected firms according to their in-degree and (separately) their out-degree, before, during, and after the U.S. Financial Crisis.  Several observations are in order.  First, Lehman Brothers (LEH) has a large out-degree in the January-June 2006 and July-December 2006 networks.  During the intervening time period (April-September 2006), it has a large in-degree.  That our methodology should identify Lehman Brothers as a highly connected firm in the lead-up to the crisis is interesting given that the broker-dealer's involvement in subprime mortgage lending has been recognized as a key contributor to the crisis [\citet{friedman2011caused}].  American International Group (AIG) is also highly connected before the crisis; in fact, it is one of the top firms according to out-degree in six of the nine networks that span 2006-2008.  It is a top firm by in-degree during the January-June 2007 period.  Like Lehman Brothers, AIG played a major role in the crisis through its use of collateralized debt obligations (CDOs) and credit default swaps (CDSs), and was bailed out by the federal government shortly after Lehman Brothers' collapse [\citet{friedman2011caused}].  

More generally, we note that the top firms are not always consistent across neighboring time periods.  For example, a firm might be highly connected during one time window, but not during the windows immediately preceding or following it.  (This is the case with T. Rowe Price (TROW), which has a large in-degree during April-September 2007, but neither a large in-degree nor a large out-degree during either of the other 2007 windows.)  A major exception is AIG, as noted above.  Our methodology highlights several additional firms that are known to have contributed to the crisis: Bear Stearns (BSC) is a top ``in-firm'' during April-September 2006 and a top ``out-firm'' in July-December 2007, The Federal National Mortgage Association (aka Fannie Mae; FNM) is a top out-firm during January-June 2007 and January-June 2008, and The Federal Home Loan Mortgage Corporation (aka Freddie Mac; FRE) is a top out-firm during April-September 2006.

\begin{figure}[!t]
    \centering
    \includegraphics[scale=0.35]{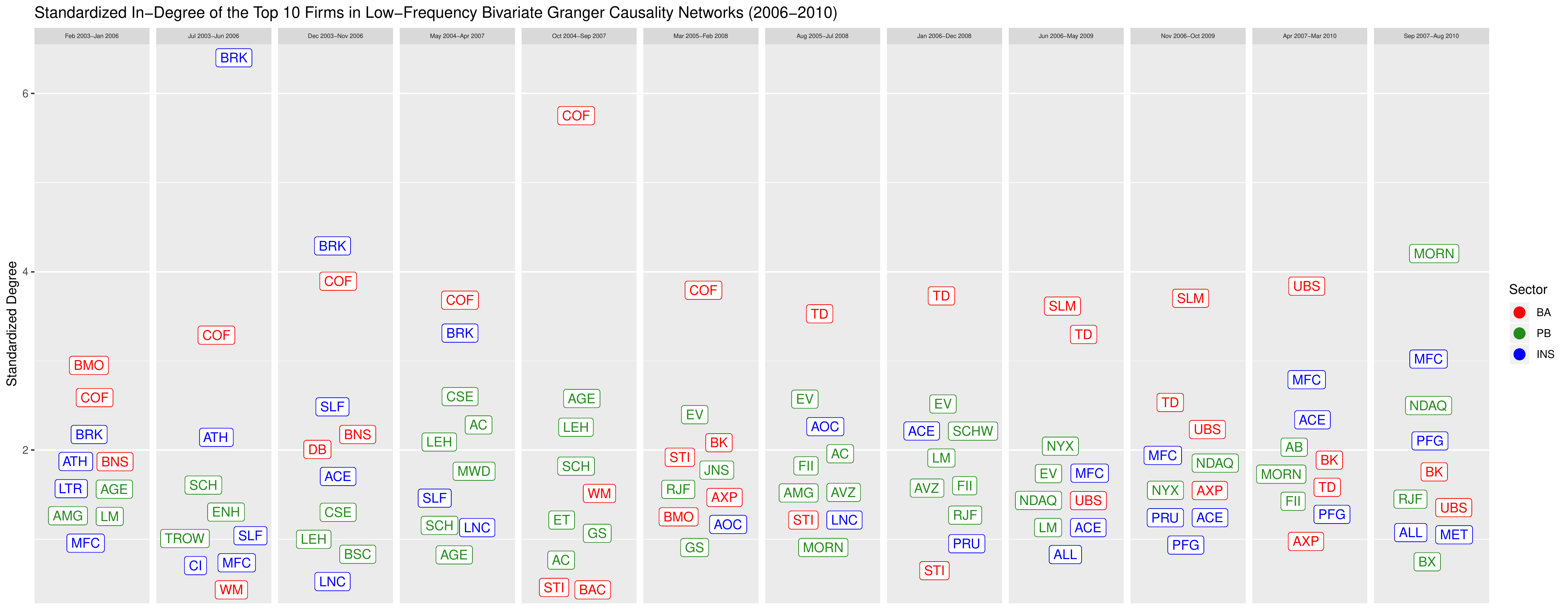}
    \includegraphics[scale=0.35]{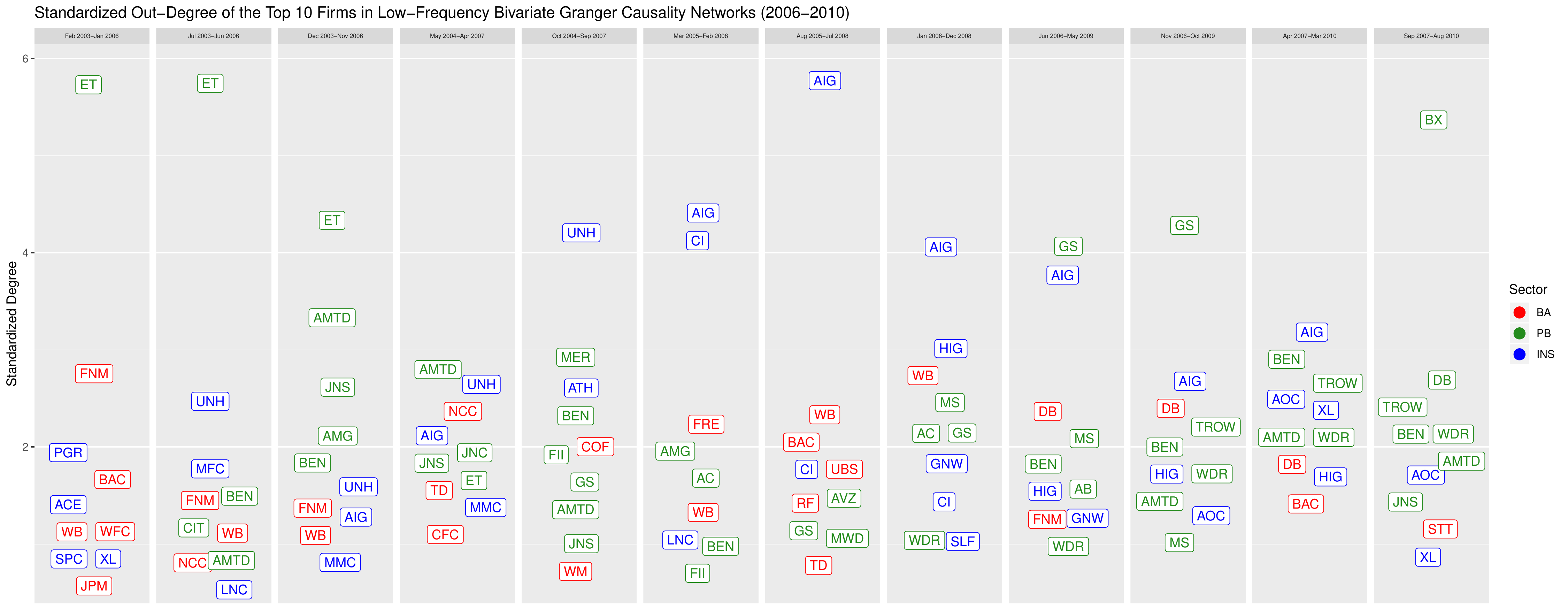}
    \caption{Highly connected firms in low-frequency bivariate Granger causality networks before, during, and after the U.S. Financial Crisis.  Firms are ranked according to their in-degree (top) and out-degree (bottom).  Standardized degrees (i.e., (firm degree - mean network degree)/standard deviation of network degree) are plotted so as to make results comparable across different networks.  Banks (resp., broker-dealers, insurance companies) are displayed in red (resp., green, blue).  Note that we add a small amount of random noise to the $(x,y)$ coordinates of each firm so that firm labels do not overlap with one another.  Full company names for all ticker symbols are provided in Table \ref{table: ML_HFT ticker symbol table}.}
    \label{fig:top_firms_low_freq}
\end{figure}

Some of these results are consistent with those observed in monthly-scale bivariate Granger causality networks (see Figure \ref{fig:top_firms_low_freq}).  In the monthly-scale networks, as in their high-frequency counterparts, AIG, Fannie Mae, and Freddie Mac all have large out-degree before and/or during the crisis.  Interestingly, AIG remains a large source of risk propagation (i.e., has high out-degree) through 2010, whereas it does not have a large out-degree in any of the high-frequency networks beyond April-September 2008.  Another key difference between the low- and high-frequency settings is the role played by Lehman Brothers and Bear Stearns.  In the high-frequency networks, Lehman Brothers, and -- to a lesser extent -- Bear Stearns, emerge as top out-firms in the lead-up to the financial crisis.  In the monthly-scale networks, on the other hand, neither of these two firms have a large out-degree, although Lehman Brothers is consistently a top in-firm (absorber of risk) in 2006 and 2007.  This difference raises the possibility that high-frequency networks may be able to identify risk propagating firms that are not highlighted in low-frequency networks.

\begin{figure}[!t]
    \centering
    \includegraphics[scale=0.35,trim={3cm 2cm 2cm 0cm},clip]{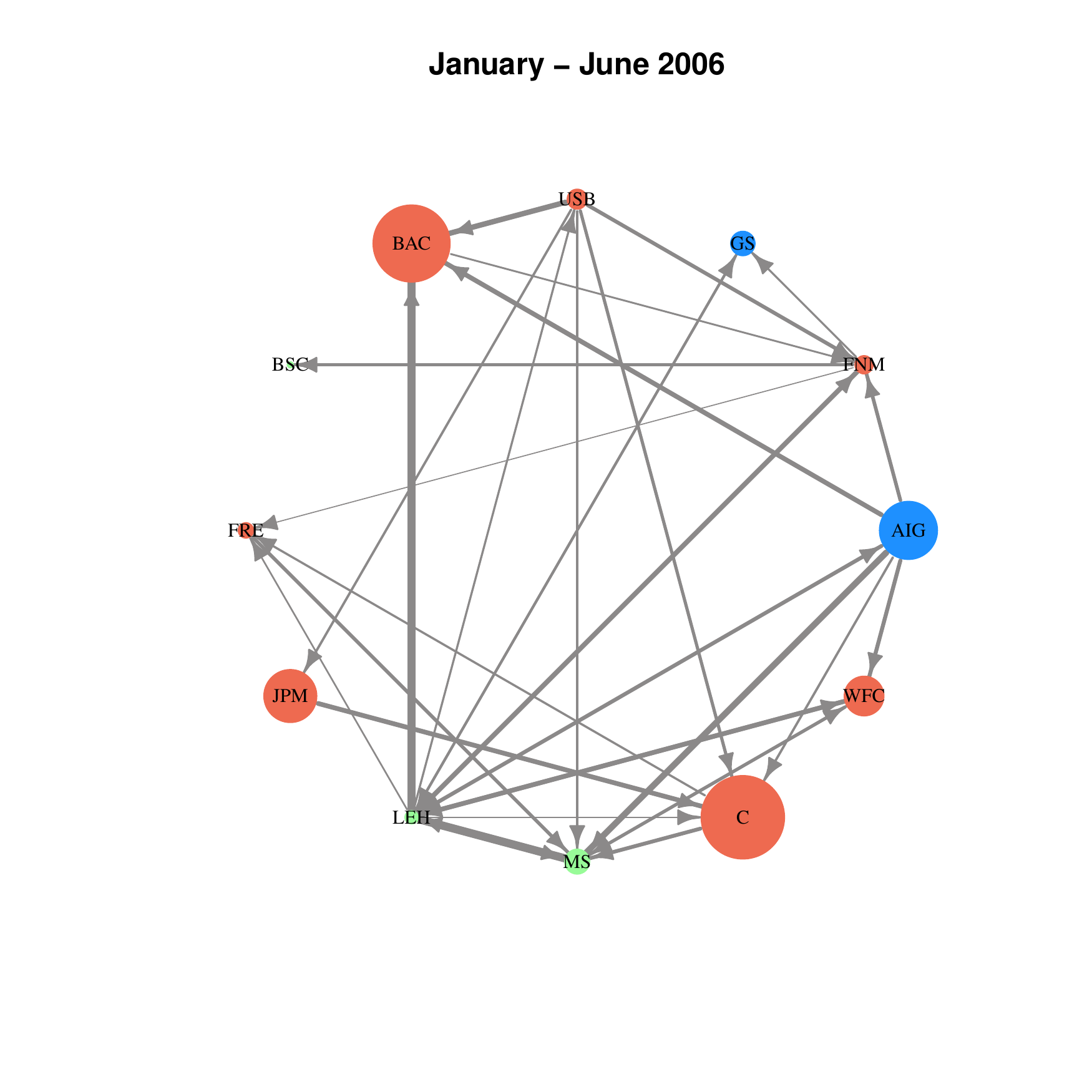}
    \includegraphics[scale=0.35,trim={3cm 2cm 2cm 0cm},clip]{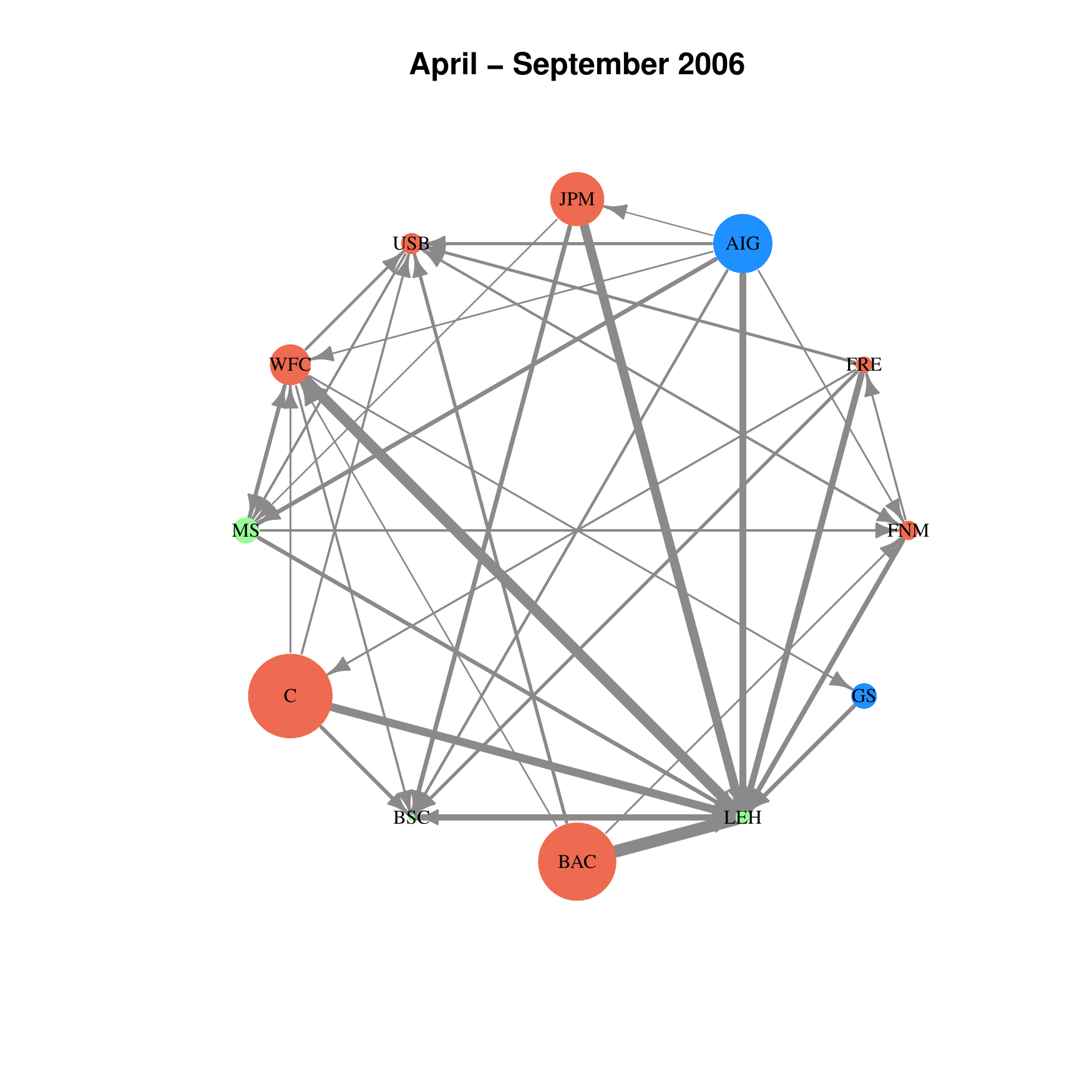}
    \includegraphics[scale=0.35,trim={3cm 2cm 2cm 0cm},clip]{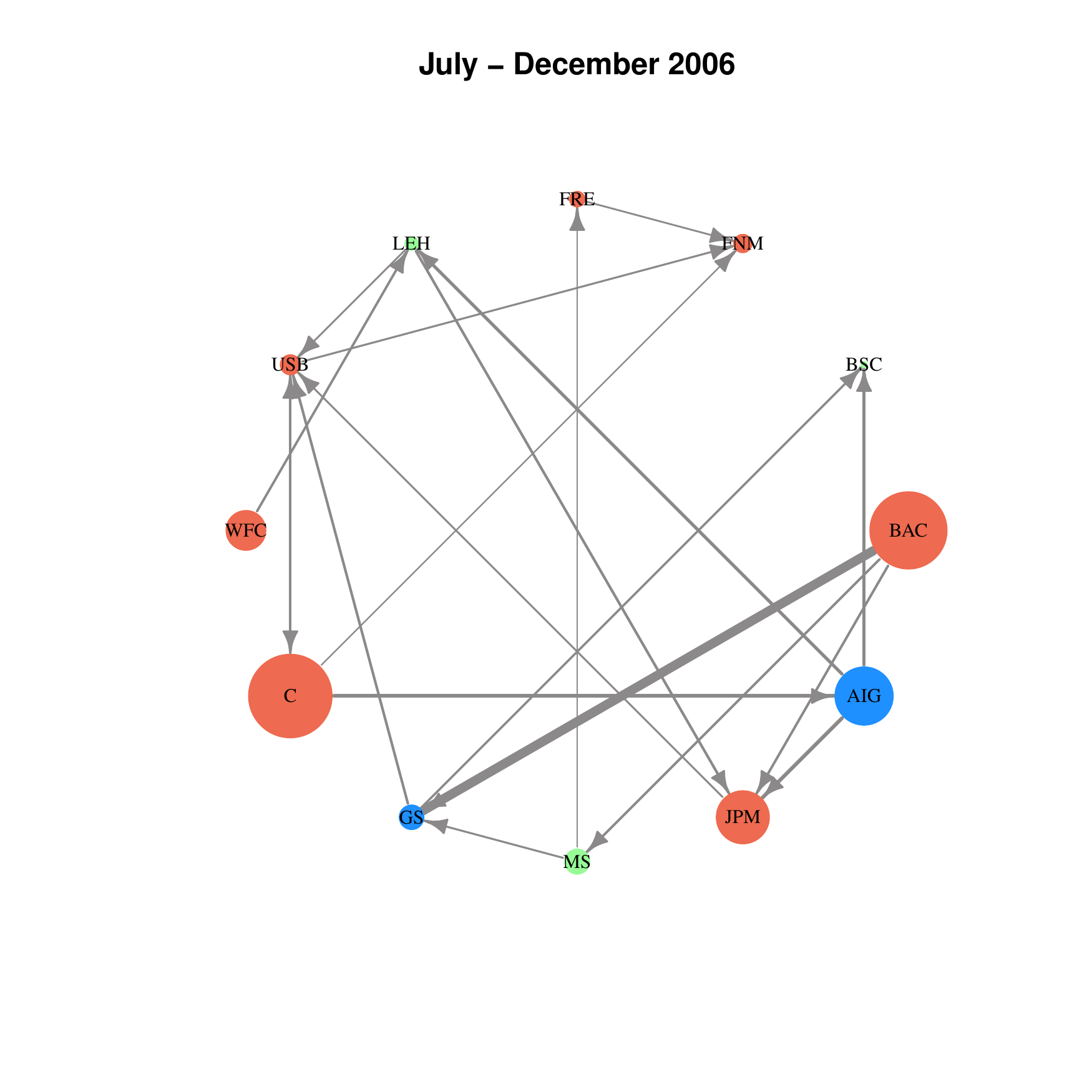}
    \includegraphics[scale=0.35,trim={3cm 2cm 2cm 0cm},clip]{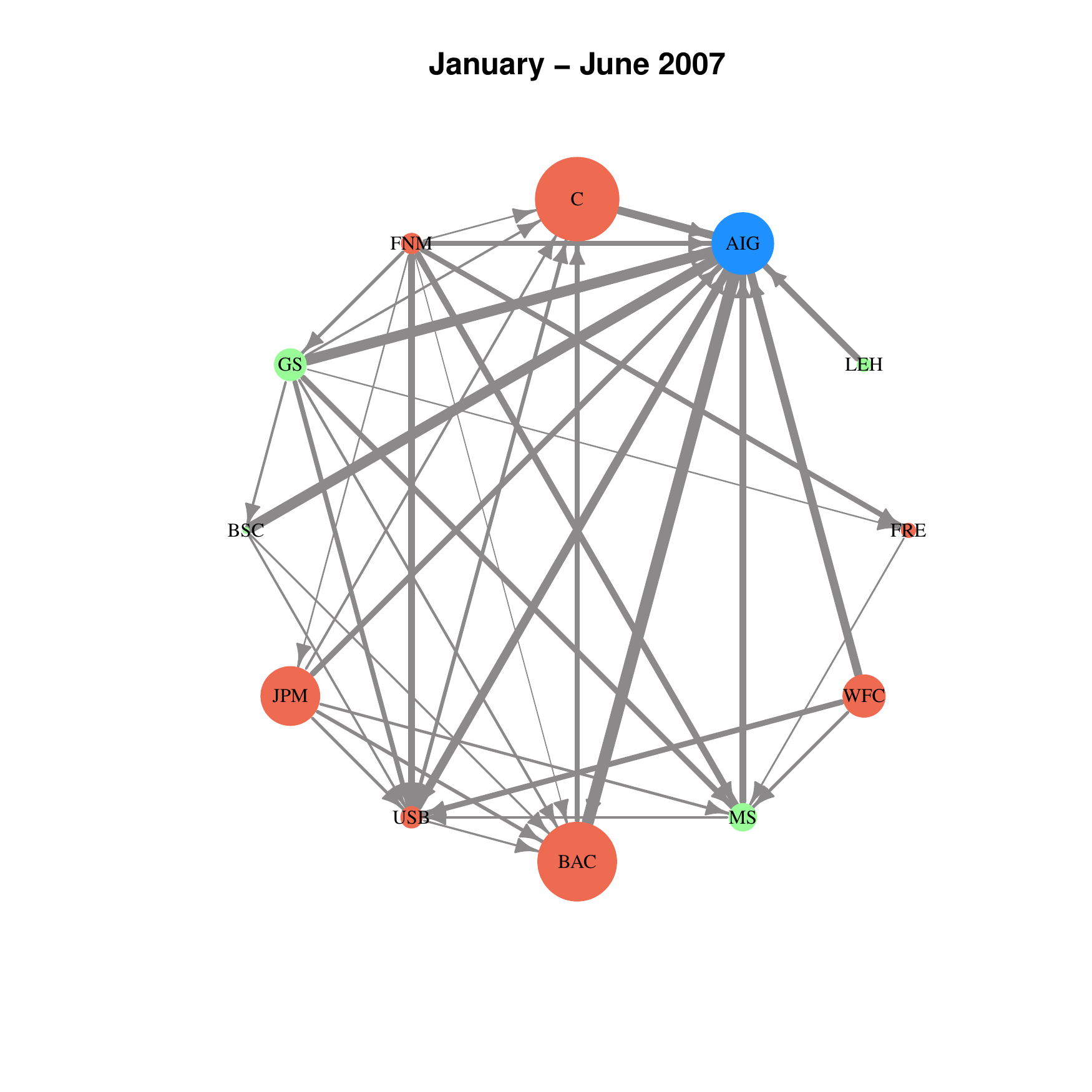}
    \includegraphics[scale=0.35,trim={3cm 2cm 2cm 0cm},clip]{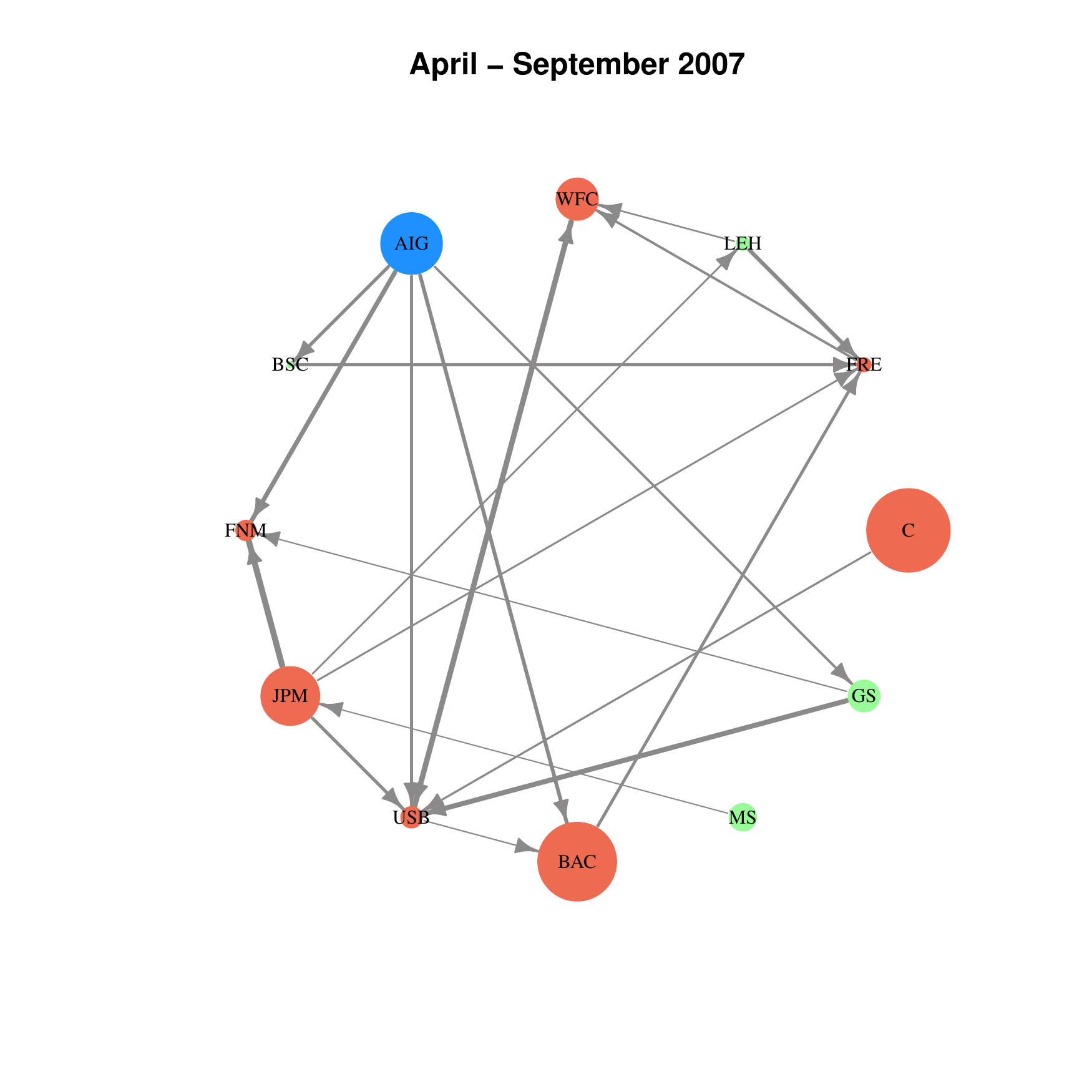}
    \includegraphics[scale=0.35,trim={3cm 2cm 2cm 0cm},clip]{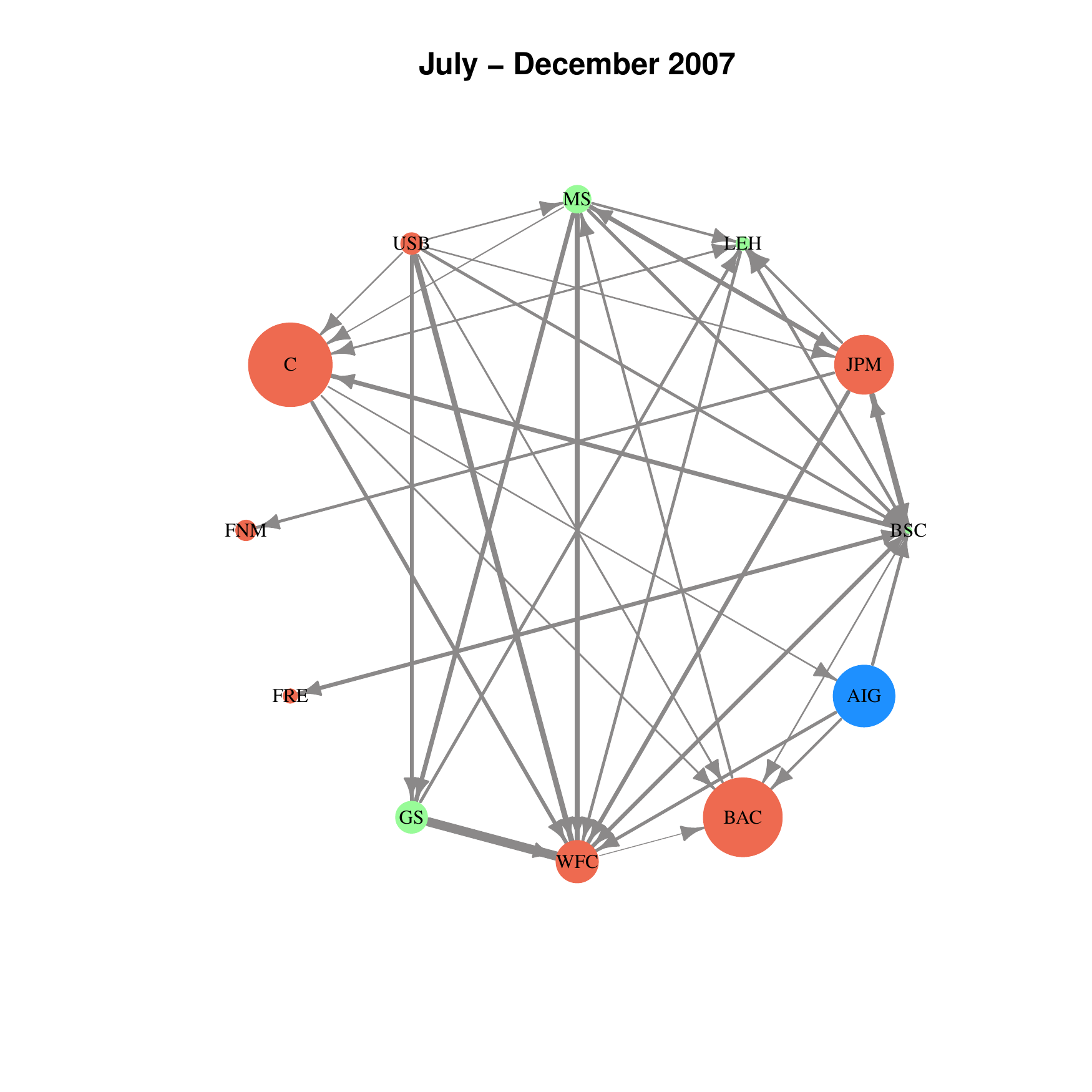}
    \caption{High-frequency realized volatility networks of financial institutions that received significant federal bailout packages (TARP funding).  Each year, the 10 institutions within our sample that received the most TARP funding are selected, in addition to Lehman Brothers and Bear Stearns.  Nodes are sized according to their market capitalization, with red (resp., green, blue) nodes indicating banks (resp., broker-dealers, insurance companies).  Edge thickness is proportional to the increase in AUC obtained by including cross-features.  Full company names for all ticker symbols are provided in Table \ref{table: ML_HFT ticker symbol table}.}
    \label{fig:subntwks_tarp_firms}
\end{figure}

To further explore the behavior of systematically important financial institutions, we consider subnetworks of firms that received considerable government assistance during or after the crisis (see Figure \ref{fig:subntwks_tarp_firms}).  The size of node $i$ is proportional to the market capitalization of firm $i$, while the thickness of edge $i \rightarrow j$ is proportional to the increase in AUC obtained by using the features of firm $i$ to predict the change in realized volatility of firm $j$.  We select -- for each year between 2006 and 2008 -- the 10 firms within our sample that received the largest amount of Troubled Asset Relief Program (TARP) funding, which was provided to companies that were deemed ``too big to fail''\footnote{A firm is considered ``too big to fail'' if its collapse would result in significant damage to the economy.} [\citet{kiel2013bailout}].  (For 2006 and 2007, we also include Lehman Brothers and Bear Stearns, which did not receive TARP funding but which were crucial firms during this period.)  

Figure \ref{fig:subntwks_tarp_firms} highlights the role played by Lehman Brothers in the lead-up to the crisis.  For example, in the April-September 2006 network, Lehman Brothers has incoming edges from all but one of the other firms and is particularly influenced by Bank of America (AUC increase = 0.305), Wells Fargo (AUC increase = 0.264), and JPMorgan Chase (AUC increase = 0.223).  In early 2007, AIG emerges as a firm having large in-degree, including from Bank of America (AUC increase = 0.297), Bear Stearns (AUC increase = 0.262), and Goldman Sachs (AUC increase = 0.259).  Bank of America also has a large in-degree.  By July-December 2007 (months before its collapse in March 2008), Bear Stearns has many incoming edges, the strongest of which is from the company that would come to purchase it, JP Morgan Chase (AUC increase = 0.144).  In 2008, several of the firms previously considered are no longer present in our sample, whether because of their collapse (e.g., Lehman Brothers, Bear Stearns\footnote{Recall that firms are only included in our sample if they are actively traded during the entire year.  Bear Stearns was acquired by JPMorgan Chase in March 2008, while Lehman Brothers filed for bankruptcy in September 2008.}) or because they are no longer among the top 75 financial institutions by market capitalization (e.g., Freddie Mac).  However, firms like Citigroup, Wells Fargo, AIG, JPMorgan Chase, and Fannie Mae continue to have large in- and/or out-degree.

\subsection{Cross-Asset Information Flow Between Small and Large Firms}
\label{sec:ML_smallLargeAnalysis}

Having constructed financial networks in Section 1.6.1, we now seek to address why there are edges between some firms and not others.  One natural hypothesis, supported by the literature, is that edges may run from large firms to small firms. Indeed \citet{lo1990contrarian} provides empirical evidence of a lead-lag relationship between the weekly returns of large and small market capitalization stocks.  In particular, the authors divide a sample of 551 stocks into size-based quintiles and form equal-weighted portfolios for each quintile.  The correlation between the returns of the first quintile's portfolio, at week $t-1$, and the returns of the fifth quintile's portfolio, at week $t$, is found to be 27.6 percent.  (No evidence is discovered of the reverse -- that is, of small stock returns leading large stock returns.)  

Various explanations for this pattern exist.  Among them is the theory that market-wide information is first absorbed into the prices of large stocks (which tend to be actively traded) and subsequently into the prices of small stocks (which tend to be less frequently traded) [\citet{brennan1993investment}].  More recently, \citet{chordia2011liquidity} performs a Granger causal analysis on value-weighted portfolios of large and small market capitalization stocks.  The authors regress daily returns of these portfolios on lagged values of returns, volatilities, and quoted spreads, and find that the returns of the large stock portfolio lead the returns of the small stock portfolio, especially when the large stocks experience low liquidity.

While \citet{chordia2011liquidity} includes a variety of financial variables (returns, volatilities, and quoted spreads) in their analysis, we are thus far unaware of any studies that examine whether \textit{market microstructure variables} yield lead-lag relationships between small and large firms.  Our random forest methodology, however, lends itself well to this question.  Indeed we can assess whether the microstructure features of small (resp., large) firms are useful in predicting future increases and decreases in market measures of large (resp., small) firms.  Notice that our method -- as previously described -- is implemented on a firm-by-firm basis; that is, we use microstructure variables of firm $y$ (and possibly of a second firm, $x$) to forecast market measures of firm $y$.  This procedure is fundamentally different from the analysis in \citet{chordia2011liquidity}, which considers financial variables that have been aggregated over firms of similar size.  So that we may compare our results to those in \citet{chordia2011liquidity}, we perform a similar aggregation.

To begin, we consider all banks, primary broker-dealers, and insurance companies that were active on each trading day of 2018.  We rank these firms according to their average monthly market capitalization and take our final set of firms to be those in the first and seventh capitalization deciles.  The first decile (54 firms) represents large stocks and the seventh decile (55 firms) represents small stocks.  We use the seventh decile because stocks in lower tiers are likely to trade so infrequently as to make missing values a problem in our downstream analysis.  Next, for each stock, we form time series of its microstructure variables and market measures, and then compute value-weighted averages for both small and large firms (separately).  For example, the aggregate time series of the Roll measure for small and large firms is given by
\begin{align}
\label{eq:aggregateRollSmall}
    R^{small}_t &= \frac{\sum_{i = 1}^{n_s} \left(MCAP^{s_i}\times R^{s_i}_t\right)}{\sum_{i = 1}^{n_s} MCAP^{s_i}}, \\
\label{eq:aggregateRollLarge}
    R^{large}_t &= \frac{\sum_{i = 1}^{n_\ell} \left(MCAP^{\ell_i}\times R^{\ell_i}_t\right)}{\sum_{i = 1}^{n_\ell} MCAP^{\ell_i}},
\end{align}
where $\left\{s_i\right\}_{1 \leq i \leq n_s}$ and $\left\{\ell_i\right\}_{1 \leq i \leq n_\ell}$ denote the set of small and large firms, respectively, $MCAP^j$ is the average monthly MCAP of firm $j$, and $R^j_t$ is the value of the Roll measure of firm $j$ at time $t$.  We form aggregate time series for Amihud's lambda, VPIN, kurtosis, and realized volatility in an analogous manner to \eqref{eq:aggregateRollSmall} and \eqref{eq:aggregateRollLarge}.\footnote{Note that we do not include two of the microstructure variables described in Section \ref{sec:ML_HFT_market_micro_variables}, namely Roll impact and Kyle's lambda.  We exclude these because they were found to have relatively low predictive ability for 2018 financial firms.}  Finally, we calculate the sign of the change in average kurtosis and realized volatility:
\begin{align*}
    &sign\left(\sigma^{small}_{t+h} - \sigma^{small}_t\right) 
    &sign\left(\sigma^{large}_{t+h} - \sigma^{large}_t\right), \\
    &sign\left(K^{small}_{t+h} - K^{small}_t\right) 
    &sign\left(K^{large}_{t+h} - K^{large}_t\right),
\end{align*}
where, as before, $h$ is a forecast horizon of 50 time bars.  There are 2,881 observations in our final dataset and we perform 10-fold cross-validation (over the entirety of 2018) to evaluate the random forest classifier's performance.

\begin{figure}[!t]
    \centering
    \includegraphics[width = 0.95\textwidth]{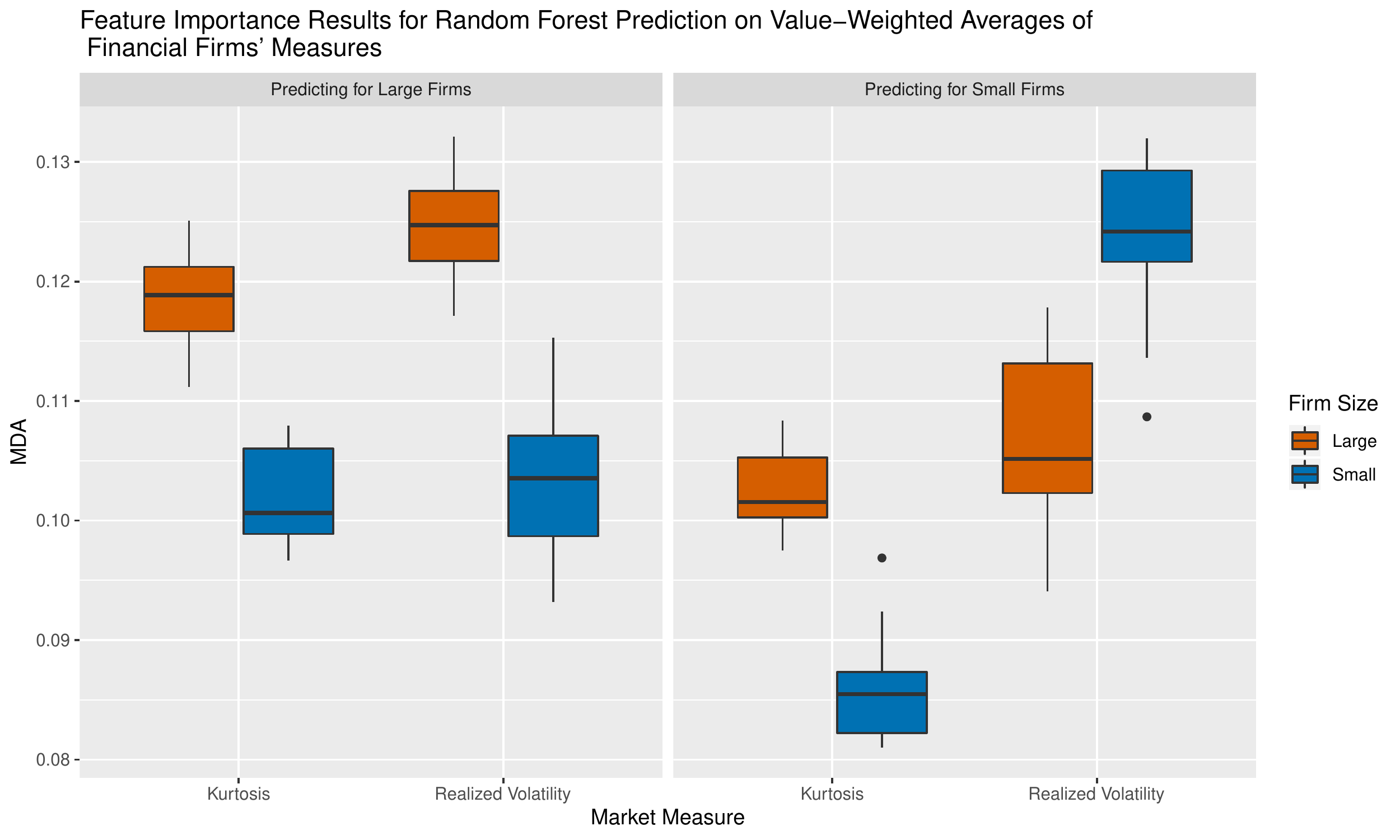}
    \caption{Distribution of the average mean decrease in accuracy (MDA), grouped by firm size.  For example, the leftmost boxplot illustrates the distribution of $\big\{\overline{M}^i_{Large}\big\}_{1 \leq i \leq 10}$, where $\overline{M}^i_{Large}$ denotes the average of the MDA values for the large firms' Roll measure, Amihud's lambda, and VPIN, evaluated over test set $i$.  The left (resp., right) panel displays feature importance results when forecasting the sign of the change in kurtosis and realized volatility for large (resp., small) financial firms.}
    \label{fig:bigSmallFinancialsFeatImp}
\end{figure}

The first question we address is which features are important for predicting market measure changes in firms of different sizes.  We consider four prediction scenarios: (i) kurtosis for large firms, (ii) realized volatility for large firms (iii) kurtosis for small firms, and (iv) realized volatility for small firms.  In each case, we include cross-features in our random forest model; that is, we use microstructure variables for both small and large firms.  Furthermore, for each test set, we compute the mean decrease in accuracy (MDA) for each feature and average the results by firm size.  Let $M^i_x$ denote the MDA of feature $x$ on test set $i$, where $i = 1, 2,..., 10$.  Then, for each $i$, we calculate
\begin{align*}
    \overline{M}^i_{Small} &= mean\left\{M^i_{Roll.Small}, M^i_{Amihud.Small}, M^i_{VPIN.Small}\right\}, \\
    \overline{M}^i_{Large} &= mean\left\{M^i_{Roll.Large}, M^i_{Amihud.Large}, M^i_{VPIN.Large}\right\}.
\end{align*}
Figure \ref{fig:bigSmallFinancialsFeatImp} displays the distributions of $\big\{\overline{M}^i_{Small}\big\}_{1 \leq i \leq 10}$ and of $\big\{\overline{M}^i_{Large}\big\}_{1 \leq i \leq 10}$ for each of the four prediction scenarios.

We observe that, under all scenarios but one, the microstructure variables of large firms are more important than those of small firms.  For example, when forecasting kurtosis for large firms, the median large firm MDA is approximately 0.12 while the median small firm MDA is closer to 0.1.  Qualitatively similar results hold when predicting realized volatility for large firms and kurtosis for small firms.  On the other hand, this pattern is reversed when we forecast realized volatility for small firms, in which case the small firms' features have higher MDA.  Interestingly, though, the distributions of $\left\{\overline{M}^i_{Large}\right\}_i$ and $\left\{\overline{M}^i_{Small}\right\}_i$ have some overlap in this case, whereas in all other scenarios, the distribution of large firm MDA values lies entirely above the distribution of small firm MDA values.  

To an extent, these results are consistent with those reported in \citet{chordia2011liquidity} and \citet{lo1990contrarian}, where the weekly and daily returns of large stocks were found to lead those of small stocks (but not the reverse).  Our analysis reveals a similar lead-lag pattern in a high-frequency setting: microstructure variables of large firms have predictive power when forecasting the sign of the change in kurtosis of the small firms' returns distribution.  Moreover, when forecasting for large firms, the microstructure features of small firms are found to be \textit{less} important than those of large firms.  This conforms with earlier findings that small firm returns do not lead large firm returns.

\begin{figure}[!t]
    \centering
    \includegraphics[angle=-90,scale=0.6, trim = {1cm 1cm 1cm 1cm}, clip]{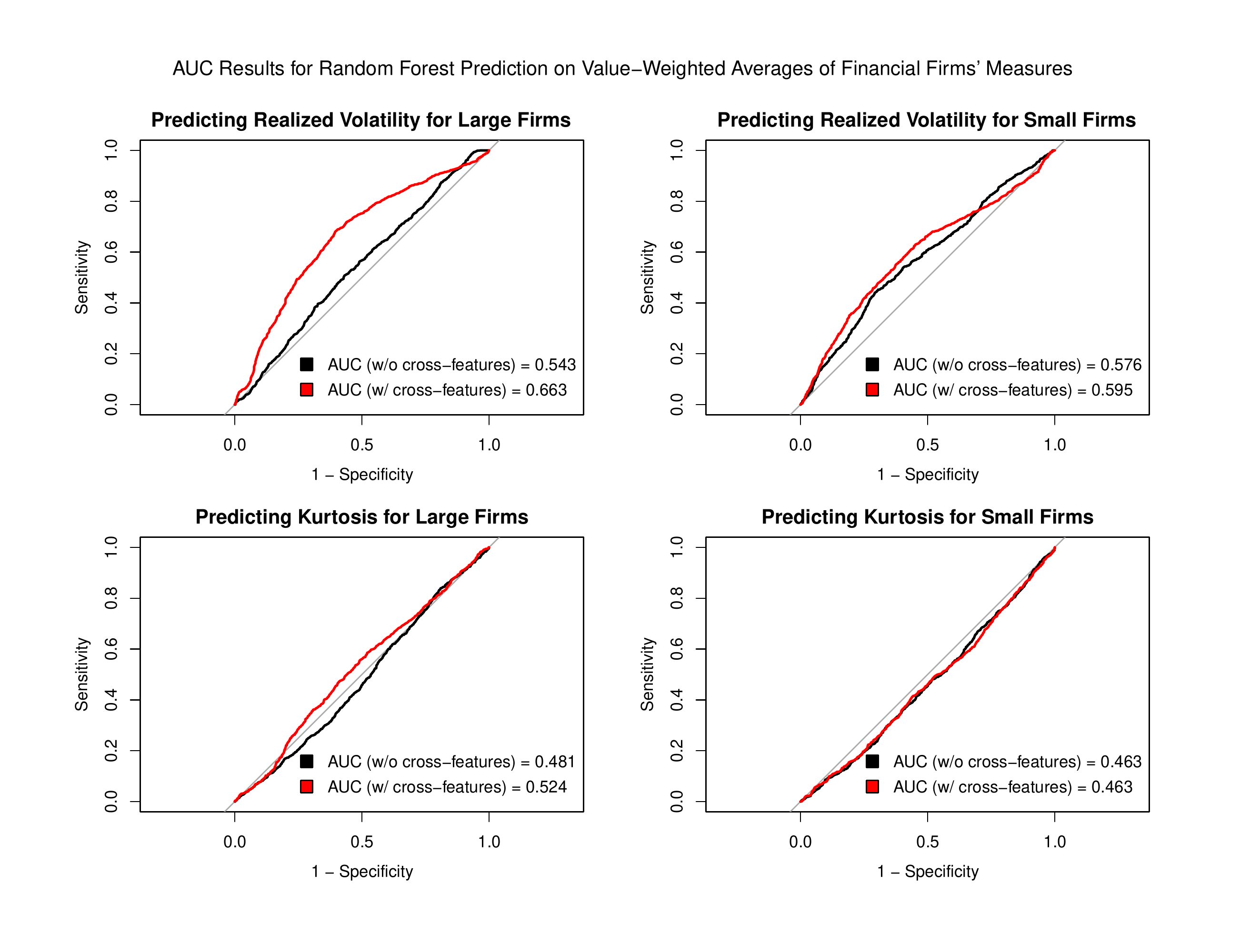}
    \caption{ROC curves for predicting the sign of the change in realized volatility (top row) and in kurtosis (bottom row) for large firms (left column) and small firms (right column).  In each case, the random forest was fit twice, once without cross-features (e.g., using only large firms' features to predict large firms' measures) and again with cross-features.  Thus, for each prediction scenario, two ROC curves are displayed: red (resp., black) curves indicate that cross-features were (resp., were not) used.  The area under the curve (AUC) is reported in the lower right corner of each plot.}
    \label{fig:bigSmallFinancialsROCCurves}
\end{figure}

We now turn our attention to the question of whether adding cross-features improves the random forest's predictive ability.  We find that the results here are mixed (see Figure \ref{fig:bigSmallFinancialsROCCurves}).  Including cross-features yields a significant increase in AUC when forecasting realized volatility, and -- to a lesser extent -- kurtosis, for large firms.  There is only minor predictive improvement, however, when using cross-features to predict realized volatility for small firms, and virtually no change when predicting kurtosis for small firms.\footnote{We note that, regardless of firm size, our model has more success in forecasting realized volatility than it does in forecasting kurtosis.}  As a robustness check, we repeat the analyses presented in this section on a set of information and communications technology (ICT) firms (rather than financial firms, as discussed here).  We find that the ICT feature importance results are qualitatively similar to those we present here, though the ROC results are not.  Our complete findings are shown in Appendix \ref{sec:ML_HFT_appendix}.
\section{Conclusion}
\label{sec:ML_HFT_discussion}

We estimate financial networks by determining whether cross-effects in \textit{intraday} trade data exist between each pairs of firms in our sample. We detect these cross-effects by assessing whether microstructure measures of one firm improve our ability to forecast the sign of the change in a market measure (either realized volatility or returns kurtosis) of another firm, where predictive performance is measured via the area under the curve (AUC). Because we learn our networks from high-frequency trade data, which tends to be both nonlinear and nonstationary, we use a random forest to forecast market measure changes. Random forests, a popular machine learning tool, provide a great deal of modeling flexibility as they do not impose a particular functional form on the data. We apply these methods to the trade data of large U.S. financial institutions, demonstrating how our networks can be used to answer the same questions posed by researchers in the low-frequency setting (e.g., how network connectivity evolves over time and which types of firms interact with one another). 

High-frequency financial networks have the potential to yield novel insights into the workings of the financial system.  Future work in this direction includes refining our network estimation procedures (e.g., by changing the microstructure variables used as features, or by considering different market measures for prediction).  There are a number of hyper-parameters in our random forest model ($W$, the length of the lookback window; $h$, the forecast horizon; the length of the time bar, etc.) and we have yet to perform an exhaustive review of how these parameters impact our final results.  

Moreover, the networks we construct are based on \textit{bivariate} analyses; that is, by testing for predictive improvements in firm $x$ when we include the features of one additional firm, $y$.  We could instead undertake a multivariate analysis wherein we include features of \textit{all} firms in order to predict the change in the market measure of firm $x$. Such an analysis would give assurance that any cross-predictability detected between firms $x$ and $y$ is indeed due to the measures of firm $y$ and not to measures of a firm that is correlated with $y$ (i.e., indirect associations).  Preliminary work in this direction has yielded mixed results; however, it is possible that by adjusting our model beyond the standard random forest, we may be able to make further progress.  On that note, it is also interesting to consider how time series models (e.g., an autoregressive integrated moving average (ARIMA) model with exogenous variables) would fare in predicting changes in market measures.

\section*{Funding}
KK and SB were supported in part by NIH award R01GM135926. SB also acknowledges partial support from NSF awards DMS-1812128, DMS-2210675 and NIH award R21NS120227.

\bibliographystyle{tfcse}
\bibliography{mlbib}

\appendix
\section{Additional Details}\label{sec:ML_HFT_appendix}

In Section \ref{sec:ML_smallLargeAnalysis}, we apply a random forest model to the aggregated market measures of different sized financial firms to assess whether firm size impacts cross-predictability (i.e., whether trade information from large (resp., small) firms improves the predictability of small (resp., large) firms' market measures).  Here we repeat our analysis on a set of information and communications technology (ICT) firms, with the goal of determining whether our results vary by industry. 

We determine ICT firms on the basis of their North American Industry Classification System (NAICS) code, which was obtained through the Center for Research in Security Prices (CRSP) database [\citet{CRSP}, \citet{NAICS}].  To begin, we select all firms having any of 10 NAICS industry codes listed in Table \ref{tab:naics_codes}.  We sort these firms according to their average market capitalization over 2018 and retain 47 firms from the first decile (representing large technology firms) and 47 from the seventh decile (representing small technology firms). 

\begin{figure}[!t]
    \centering
    \includegraphics[width = 0.95\textwidth]{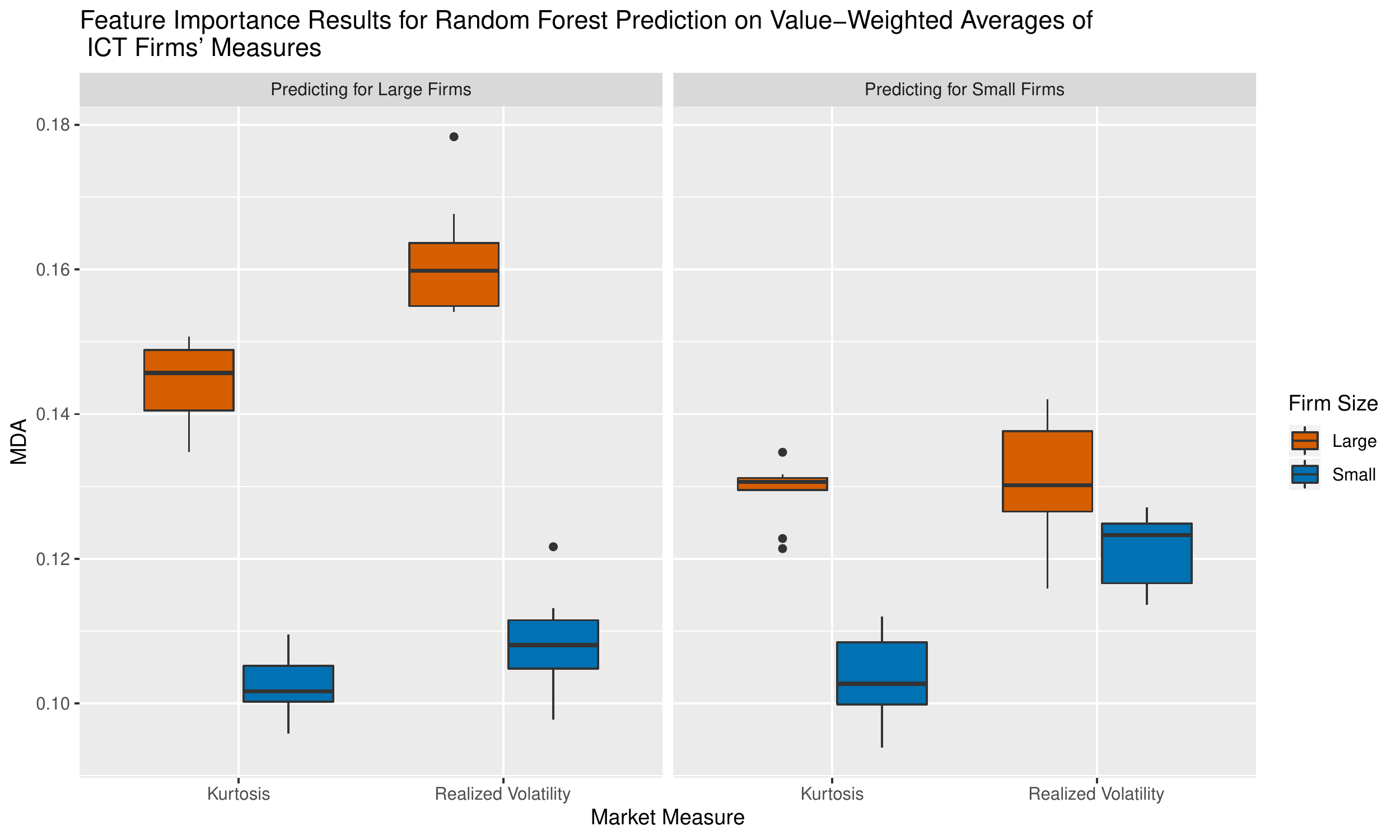}
    \caption{Distribution of the average MDA, grouped by firm size.  The left (resp., right) panel displays feature importance results when forecasting the sign of the change in kurtosis and realized volatility for large (resp., small) ICT firms.}
    \label{fig:bigSmallFeatImps_Tech}
\end{figure}

\begin{figure}[!t]
    \centering
    \includegraphics[angle=-90,width = 0.95\textwidth, trim = {1cm 1cm 1cm 1cm}, clip]{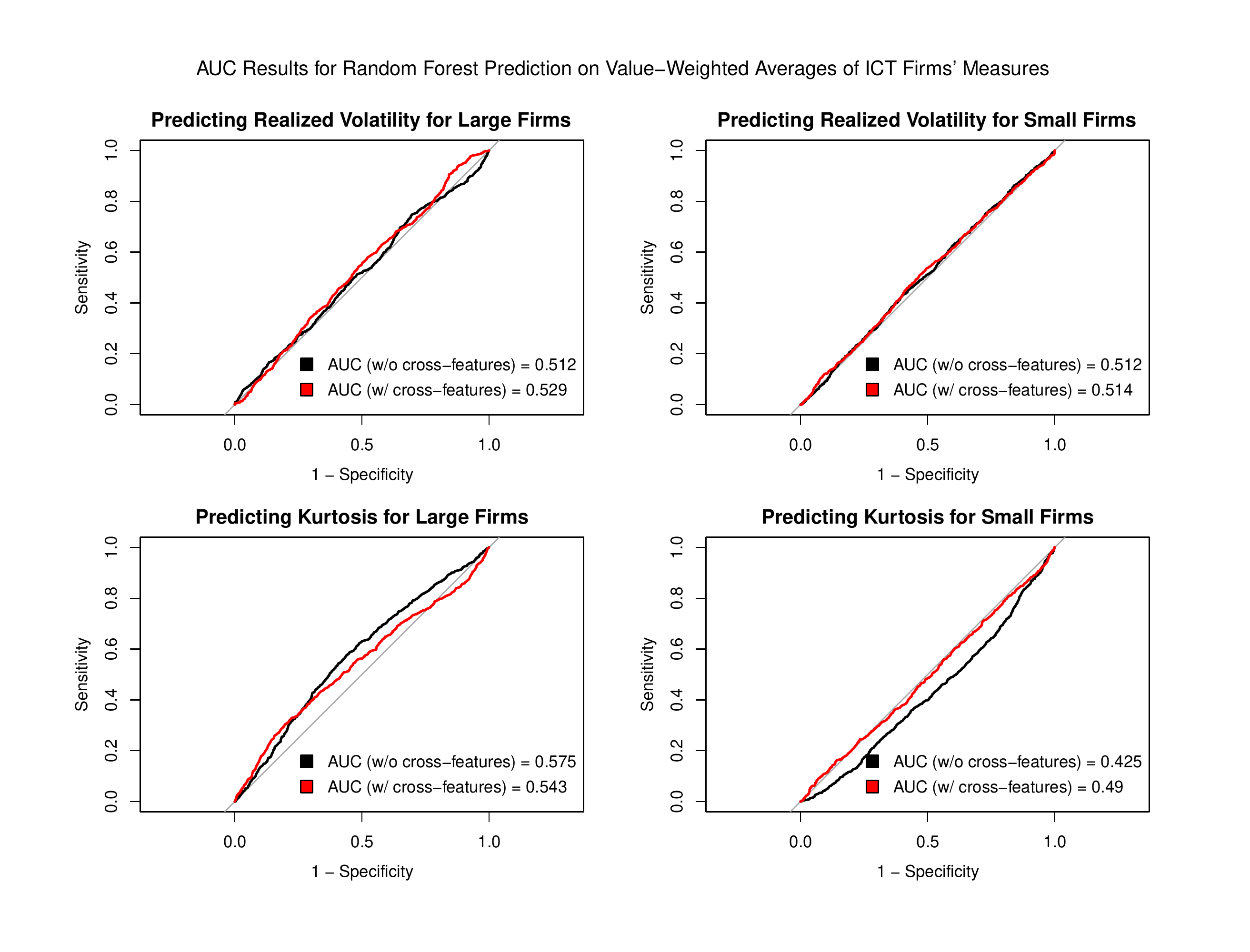}
    \caption{ROC curves for predicting the sign of the change in realized volatility (top row) and in kurtosis (bottom row) for large (left column) and small (right column) ICT firms.  For each prediction scenario, two ROC curves are displayed: red (resp., black) curves indicate that cross-features were (resp., were not) used.  The area under the curve (AUC) is reported in the lower right corner of each plot.}
    \label{fig:bigSmallROCCurves_Tech}
\end{figure}

Figure \ref{fig:bigSmallFeatImps_Tech} displays MDA feature importance results.  On average, large firms' features are more important than small firms' features, regardless of whether we are forecasting realized volatility or kurtosis, for large firms or for small firms.  However, the difference in feature importances is larger when predicting for large firms (left panel) than for small firms (right panel), which suggests (a) that small firms carry little information about large firms, and (b) that large firms do contain some information about small firms, but the small firm features are still significant.  These results are qualitatively similar to what we obtain for financial firms, except that, for the latter, small firms' features are more important than large firms' when predicting small firm realized volatility.  In Figure \ref{fig:bigSmallROCCurves_Tech}, we show that including cross-features in the random forest model yields very little change in the AUC.  An exception is when we predict kurtosis for small firms (bottom right plot), in which case we see an appreciable increase in AUC when we add large firms' features. 

\begin{table}[H]
    \centering
    \begin{tabular}{l|l}
        \textbf{NAICS Code} & \textbf{Description}  \\ \hline\hline
         3341 & Computer and peripheral equipment manufacturing \\ \hline
         3342 & Communications equipment manufacturing \\ \hline
         3344 & Semiconductor and other electronic component manufacturing \\ \hline
         3345 & \parbox{10cm}{Navigational, measuring, electromedical, and control instruments manufacturing} \\ \hline
         5112 & Software publishers \\ \hline
         5161 & Internet publishing and broadcasting \\ \hline
         5179 & Other telecommunications \\ \hline
         5181 & Internet service providers and Web search portals \\ \hline
         5182 & Data processing, hosting, and related services \\ \hline
         5415 & Computer systems design and related services \\
    \end{tabular}
    \caption{North American Industry Classification System (NAICS) industry codes for information and communications technology firms.}
    \label{tab:naics_codes}
\end{table}

\begin{longtable}[H]{c | c | c}
\textbf{Firm Name} & \textbf{Sector} & \textbf{Ticker Symbol} \\ \hline\hline
ALLIANCEBERNSTEIN HLDG & PB & AB, AC \\
ACE LTD & INS & ACE \\
AETNA INC NEW & INS & AET \\
A F L A C INC & INS & AFL \\
AMERICAN INTERNATIONAL GROUP & INS & AIG \\
APOLLO INVESTMENT CORP & PB & AINV \\
ASSURANT INC & INS & AIZ \\
ALLSTATE CORP & INS & ALL  \\
AFFILIATED MANAGERS GROUP INC & PB & AMG \\
AMERIPRISE FINANCIAL INC & PB & AMP \\
AMERITRADE HOLDING CORP NEW & PB & AMTD \\
AON CORP & INS & AOC, AON \\
AMERICAN EXPRESS CO & BA & AXP \\
BANK OF AMERICA CORP & BA & BAC \\
B B \& T CORP & BA & BBT \\
FRANKLIN RESOURCES INC & PB & BEN \\
BANK NEW YORK INC & BA & BK \\
BLACKROCK INC & PB & BLK \\
BANK MONTREAL QUE & BA & BMO \\
BANK OF NOVA SCOTIA & BA & BNS \\
C B O T HOLDINGS INC & PB & BOT \\
BEAR STEARNS COMPANIES INC & PB & BSC \\
BLACKSTONE GROUP L P & PB & BX \\
CITIGROUP & BA & C \\
CHUBB CORP & INS & CB \\
COUNTRYWIDE FINANCIAL CORP & BA & CFC \\
C I G N A CORP & INS & CI \\
CIT GROUP & PB & CIT \\
CANADIAN IMPERIAL BANK COMMERCE & BA & CM \\
CHICAGO MERCANTILE EXCH HLDG INC & PB & CME \\
C N A FINANCIAL CORP & INS & CNA \\
CAPITAL ONE FINANCIAL CORP & BA & COF \\
COVENTRY HEALTH CARE INC & INS & CVH \\
DEUTSCHE BANK A G & BA & DB \\
DISCOVER FINANCIAL SERVICES & BA & DFS \\
E TRADE FINANCIAL CORP & PB & ET, ETFC \\
EATON VANCE CORP & PB & EV \\
FEDERATED INVESTORS INC PA & PB & FII \\
FEDERAL NATIONAL MORTGAGE ASSN & BA & FNM \\
FEDERAL HOME LOAN MORTGAGE CORP & BA & FRE \\
GREENHILL \& CO INC & PB & GHL \\
GENWORTH FINANCIAL INC & INS & GNW \\
HARTFORD FINANCIAL SVCS GRP INC & PB & HIG \\
BLOCK H \& R INC & BA & HRB \\
HUMANA INC & INS & HUM \\
INTERACTIVE DATA CORP & PB & IDC \\
INVESCO LTD & PB & IVZ \\
JEFFERIES GROUP INC NEW & PB & JEF \\
NUVEEN INVESTMENTS INC & PB & JNC \\
JANUS CAP GROUP INC & PB & JNS \\
JPMORGAN CHASE \& CO & BA & JPM \\
LAZARD LTD & PB & LAZ \\
LEHMAN BROTHERS HOLDINGS INC & PB & LEH \\
LEGG MASON INC & PB & LM \\
LINCOLN NATIONAL CORP IN & INS & LNC \\
MERRILL LYNCH \& CO INC & PB & MER \\
METLIFE INC & INS & MET \\
MANULIFE FINANCIAL CORP & INS & MFC \\
MARSH \& MCLENNAN COS INC & INS & MMC \\
MORNINGSTAR INC & PB & MORN \\
MORGAN STANLEY DEAN WITTER \& CO & PB & MS \\
M \& T BANK CORP & BA & MTB \\
NATIONAL CITY CORP & BA & NCC \\
NASDAQ STOCK MARKET INC & PB & NDAQ \\
NYMEX HOLDINGS INC & PB & NMX \\
NORTHERN TRUST CORP & BA & NTRS \\
N Y S E GROUP INC & PB & NYX \\
PEOPLES UNITED FINANCIAL INC & BA & PBCT \\
PRINCIPAL FINANCIAL GROUP INC & INS & PFG \\
PROGRESSIVE CORP OH & INS & PGR \\
P N C FINANCIAL SERVICES GRP INC & BA & PNC \\
PARTNERRE LTD & INS & PRE \\
PRUDENTIAL FINANCIAL INC & INS & PRU \\
EVEREST RE GROUP LTD & INS & RE \\
REGIONS FINANCIAL CORP & BA & RF \\
RAYMOND JAMES FINANCIAL INC & PB & RJF \\
ROYAL BANK CANADA MONTREAL QUE & BA & RY \\
SCHWAB CHARLES CORP NEW & PB & SCHW \\
S E I INVESTMENTS COMPANY & PB & SEIC \\
SUN LIFE FINANCIAL INC & INS & SLF \\
SLM CORP & BA & SLM \\
ST PAUL TRAVELERS COS INC & INS & STA \\
SUNTRUST BANKS INC & BA & STI \\
STATE STREET CORP & BA & STT \\
TORONTO DOMINION BANK ONT & BA & TD \\
T ROWE PRICE GROUP INC & PB & TROW \\
TRAVELERS GROUP INC & INS & TRV \\
U B S AG & BA & UBS \\
UNITEDHEALTH GROUP INC & INS & UNH \\
UNUMPROVIDENT CORP & INS & UNM \\
U S BANCORP DEL & BA & USB \\
VISA INC & BA & V \\
WACHOVIA CORP 2ND NEW & BA & WB \\
WELLS FARGO \& CO NEW & BA & WFC \\
WASHINGTON MUTUAL INC & BA & WM \\
WILLIS GROUP HOLDINGS PUB LTD CO & INS & WSH \\
X L CAPITAL LTD & INS & XL \\
\caption{Firm names, sectors, and ticker symbols.  BA: bank, PB: broker/dealer, INS: insurance.}
    \label{table: ML_HFT ticker symbol table}
\end{longtable}

\end{document}